\documentclass[manuscript,screen]{acmart}

\usepackage[utf8]{inputenc}
\usepackage{graphicx} 
\usepackage{enumitem}
\usepackage{calc}
\usepackage{multicol} 
\usepackage{listings}
\usepackage{adjustbox}
\usepackage{svg}
\usepackage{booktabs}
\usepackage{tabularx}
\usepackage{soul}
\usepackage{tabularx}
\usepackage[noend]{algpseudocode}
\usepackage{booktabs}
\usepackage{multirow}
\usepackage{graphicx}
\usepackage{textcomp}
\usepackage{tikz}
\usepackage{xcolor}
\usepackage{enumitem}
\usepackage{caption}
\usepackage[linesnumbered,ruled,lined]{algorithm2e}
\usepackage{alltt}
\usepackage{multirow}
\usepackage{color, soul}
\usepackage{textcomp}
\usepackage{framed}
\usepackage{hhline}
\usepackage{subcaption}
\usepackage[most]{tcolorbox}
\usepackage{float}
\usepackage{array}
\usepackage[utf8]{inputenc}
\usepackage{flushend}
\usepackage{makecell}
\usepackage{balance}
\usepackage{centernot}
\usepackage[T1]{fontenc}
\usepackage{amsthm}

\usepackage{stmaryrd}
\usepackage[colorinlistoftodos,textwidth=15mm,textsize=tiny,obeyFinal]{todonotes}
\usepackage{lstautogobble}

\usepackage{fancyvrb}
\usepackage{hyperref}

\definecolor{findingtitlebg}{HTML}{F2F2F2}

\newtcolorbox[
    auto counter
]{findingbox}[1][]{
    enhanced,
    unbreakable,
    colback=white,
    colframe=black!75!white,
    colbacktitle=findingtitlebg,
    coltitle=black,
    boxrule=0.6pt,
    arc=1pt,
    left=5pt,
    right=5pt,
    top=5pt,
    bottom=5pt,
    before=\par\medskip\noindent,
    after=\par\medskip,
    fonttitle=\bfseries,
    title={Finding~\thetcbcounter},
    #1
}

\definecolor{azure(colorwheel)}{rgb}{0.0, 0.5, 1.0}

\newtcbox{\mybox}[1][breakable]{on line, enlarge top by=10pt, enlarge bottom by=10pt,
     boxsep=8pt, boxrule=2pt, size=small, arc=1mm}

\definecolor{grey}{rgb}{0.7,0.7,0.7}

\newcommand{\lstbg}[3][0pt]{{\fboxsep#1\colorbox{#2}{\strut #3}}}
\lstdefinelanguage{diff}{
  basicstyle=\ttfamily\scriptsize,,
  morecomment=[f][\lstbg{red!20}]-,
  morecomment=[f][\lstbg{green!20}]+,
  morecomment=[f][\lstbg{yellow!20}]++,
  morecomment=[f][\textit]{@@},
  texcl=false
}

\definecolor{todocolor}{rgb}{0.9,0.1,0.1}
\definecolor{indiagreen}{rgb}{0.07, 0.53, 0.03}
\definecolor{hycolor}{rgb}{0.7,0.7,0.3}
\definecolor{darkbrown}{rgb}{0.4, 0.26, 0.13}

\tikzstyle{highlighter} = [
  yellow,
  line width = \baselineskip,
]

\newcounter{highlight}[page]

\AtBeginShipout{\AtBeginShipoutUpperLeft{\ifthenelse{\value{highlight} > 0}{\tikz[remember picture, overlay]{\foreach \stroke in {1,...,\arabic{highlight}} \draw[highlighter] (begin highlight \stroke) -- (end highlight \stroke);}}{}}}

\newcommand{\ignore}[1]{}
\definecolor{main-color}{rgb}{0.6627, 0.7176, 0.7764}
\definecolor{string-color}{rgb}{0.3333, 0.5254, 0.345}
\definecolor{key-color}{rgb}{0.8, 0.47, 0.196}
\lstdefinestyle{mystyle} {
    language = Java,
    basicstyle = {\ttfamily \color{main-color}},
    stringstyle = {\color{string-color}},
    keywordstyle = {\color{key-color}},
    keywordstyle = [2]{\color{lime}},
    keywordstyle = [3]{\color{yellow}},
    keywordstyle = [4]{\color{teal}},
    morekeywords = [3]{<<, >>},
    morekeywords = [4]{++},
    basicstyle=\ttfamily\scriptsize,
    commentstyle=\color{blue}\ttfamily,
    morecomment=[f][\lstbg{red!20}]-,
    morecomment=[f][\lstbg{green!20}]+,
    morecomment=[f][\lstbg{yellow!20}]++,
    morecomment=[f][\lstbg{yellow!20}]--,
    morecomment=[f][\textit]{@@},
    breaklines=true,
    texcl=false
}
\definecolor{grey}{rgb}{0.7,0.7,0.7}

\definecolor{todocolor}{rgb}{0.9,0.1,0.1}
\definecolor{indiagreen}{rgb}{0.07, 0.53, 0.03}
\definecolor{hycolor}{rgb}{0.7,0.7,0.3}
\definecolor{darkbrown}{rgb}{0.4, 0.26, 0.13}

\tikzstyle{highlighter} = [
  yellow,
  line width = \baselineskip,
]

\definecolor{main-color}{rgb}{0.6627, 0.7176, 0.7764}
\definecolor{string-color}{rgb}{0.3333, 0.5254, 0.345}
\definecolor{key-color}{rgb}{0.8, 0.47, 0.196}
\lstdefinestyle{mystyle} {
    language = Java,
    basicstyle = {\ttfamily \color{main-color}},
    stringstyle = {\color{string-color}},
    keywordstyle = {\color{key-color}},
    keywordstyle = [2]{\color{lime}},
    keywordstyle = [3]{\color{yellow}},
    keywordstyle = [4]{\color{teal}},
    morekeywords = [3]{<<, >>},
    morekeywords = [4]{++},
    basicstyle=\ttfamily\scriptsize,
    commentstyle=\color{blue}\ttfamily,
    morecomment=[f][\lstbg{red!20}]-,
    morecomment=[f][\lstbg{green!20}]+,
    morecomment=[f][\lstbg{yellow!20}]++,
    morecomment=[f][\lstbg{yellow!20}]--,
    morecomment=[f][\textit]{@@},
    breaklines=true,
    texcl=false
}
\newcommand{\microbenchmark}{Micro-benchmark}
\newcommand{\practicalbenchmark}{Practical-benchmark}

\newcommand{\rqthree}{RQ3}
\newcommand{\rqfour}{RQ4}

\newcommand{\micrototalnumber}{68}
\newcommand{\microeligiblenumber}{64}

\begin{document}

\title[Understanding the Energy Impact of Software Refactoring]{Understanding the Energy Impact of Software Refactoring: A Workload-Aware Study of Controlled Examples and Real-World Commits}







\author{Haibo Wang}
\affiliation{
  \institution{Concordia University}
  \city{Montréal}
  \state{Quebec}
  \country{Canada}
  }
\email{haibo.wang@mail.concordia.ca}

\author{Heng Li}
\affiliation{
  \institution{Polytechnique Montréal}
  \city{Montréal}
  \state{Quebec}
  \country{Canada}
  }
\email{heng.li@polymtl.ca}

\author{Shin Hwei Tan}
\authornote{Corresponding author.}
\affiliation{
  \institution{Concordia University}
  \city{Montréal}
  \state{Quebec}
  \country{Canada}
  }
\email{shinhwei.tan@concordia.ca}





\begin{abstract}
Refactoring improves software maintainability while preserving functional behavior, yet behavior preservation does not imply energy neutrality. Existing studies primarily examine isolated refactorings under fixed or simple workloads, leaving the effects of workload variation, real-world refactoring practices, explanatory factors, and energy regression identification insufficiently understood. We present the first large-scale empirical study of the energy impact of refactoring across two complementary Java benchmarks: a \textit{\microbenchmark{}}, comprising 68 refactoring types evaluated under diverse workloads, and a \textit{\practicalbenchmark{}}, containing 481 real-world refactoring commits from 430 GitHub projects. Using repeated paired energy measurements, 
we analyze workload sensitivity, refactoring patterns, explanatory factors, and the effectiveness of metric- and LLM-based regression identification.
In the \textit{\microbenchmark{}}, 199 of 384 refactoring-workload pairs (51.8\%) exhibit statistically significant energy differences, and 45.3\% of refactoring instances change energy-impact classification across workloads. 
In the \textit{\practicalbenchmark{}}, 
only 36 commits (7.5\%) show significant energy changes,
although two-thirds differ by at least 10\%. 
Refactoring type alone is insufficient to predict energy outcomes, while certain recurring refactoring combinations are associated with energy reductions.
Changes in execution time consistently explain energy variation in the controlled benchmark but correlate weakly with energy changes in real-world commits. 
Our findings highlight the need for workload-diverse evaluation of the energy impact of refactoring;  
neither existing metric-based approaches nor LLM-based predictors can reliably identify refactoring-induced energy regressions, motivating the development of more accurate techniques for predicting the energy impact of refactoring. 

\end{abstract}




\maketitle

\section{Introduction}
\label{sec:introduction}

Refactoring is one of the most prevalent forms of software evolution~\cite{1265817,10.1145/3408302}. Developers refactor code to improve its internal structure, reduce duplication, simplify comprehension, and lower future maintenance costs, while preserving the externally observable behavior of the program~\cite{fowler2018refactoring,10.1145/3715730,10.1145/2950290.2950305}. 
Unlike functional changes, refactoring aims to modify how software is implemented rather than what it does, making behavioural preservation a central requirement~\cite{10.1145/3747289,11052804}. However, preserving functional behaviour does not guarantee that other software properties remain unchanged. Refactoring can modify execution paths, object creation patterns, library interactions, exception handling, synchronization mechanisms, and compiler optimization opportunities, thereby affecting runtime characteristics. Such changes may influence not only execution time but also memory allocation, garbage collection, CPU utilisation, and ultimately energy consumption~\cite{10.1145/3485136,10.1145/2652524.2652538}. Thus, behavioural preservation does not imply energy neutrality.



Energy consumption has emerged as an important quality concern for modern software systems. While traditionally viewed as a hardware or deployment-level issue, energy efficiency is also influenced by software design and evolution decisions.
Software now operates across a wide range of environments, from battery-constrained mobile and embedded devices, where energy consumption directly affects usability and service lifetime, to cloud platforms, data centers, and continuous integration infrastructures, where large-scale computation contributes to operational costs and environmental impacts~\cite{10.1145/2307636.2307661,10.1145/2168836.2168841,10643924,11500151,verdecchia2023systematic,bolon2024review}. Empirical studies have demonstrated that software-level changes can introduce measurable energy consequences. For example, analyses of Android applications have linked energy consumption to application-level implementation choices and API usage patterns, while large-scale studies of mobile software evolution have shown that feature additions, bug fixes, and other code changes can alter energy behavior~\cite{rua2024large,10.1145/2597073.2597085}. These findings indicate that energy efficiency should be considered throughout the software lifecycle, rather than only during hardware selection or system deployment. 
Among software evolution activities, refactoring is particularly important because it is routinely applied to improve internal structure, maintainability, and long-term sustainability while preserving externally observable behavior. However, behavioral equivalence does not necessarily guarantee energy neutrality: changes intended solely to improve code quality may alter execution characteristics and energy consumption. Despite the widespread adoption of refactoring, its potential energy implications remain insufficiently understood. Understanding when and why refactoring affects energy consumption is therefore essential for enabling developers to perform energy-aware software maintenance and evolution.

Several studies have investigated whether code refactoring can change software energy consumption.~\citet{10.1145/2652524.2652538} conducted an empirical study of six commonly used refactoring types on nine Java applications. They applied each refactoring type in Java applications and compared the energy consumption of the original and refactored versions, and showed that refactoring can statistically significantly affect energy usage in both directions: some refactoring reduce energy consumption, while others increase it. They also found that easily collectible signals such as execution time and dynamic execution counts are insufficient to fully predict the energy impact of refactoring. ~\citet{park2014investigation} investigated the energy consumption of 63 refactoring types in the refactoring book~\cite{fowler2018refactoring} by constructing original and refactored examples and estimating their energy consumption by executing simple tests. Their results classified refactoring techniques into positive, neutral, and negative groups from the perspective of energy efficiency. More recently,~\citet{csanlialp2022energy} studied the energy efficiency of triple combinations of five types of refactoring. Together, these studies provide important evidence that refactoring can influence energy consumption. However, existing studies on the energy impact of refactoring share the following limitations:

\noindent\textbf{Limitation 1: The energy measurement is performed under fixed or simple workloads.} Energy measurement results can depend heavily on the workload used during execution. Prior performance studies have shown that workload variation can substantially change measured performance behavior~\cite{10172849}. This concern also applies to refactoring-energy analysis: the same refactoring may show different energy effects when the refactored code is executed with different workloads. Therefore, workload design is not merely an implementation detail, but an important factor in measuring the energy impact of refactoring.

Existing refactoring-energy studies usually rely on simple workloads. For example, \citet{park2014investigation} constructed 63 small examples for Fowler refactoring types~\cite{fowler2018refactoring} and customized them to run without external input, so that input variation would not affect power estimation. While this design improves controllability, it measures each refactoring under a fixed workload. Similarly, when functional tests are used as workloads, they are primarily designed to validate program behavior, not to expose energy-sensitive runtime behavior. Such tests may execute the refactored code only with small or typical inputs, and may not stress execution paths where the refactoring changes allocation, garbage collection, or CPU utilization. To address this limitation, we construct a workload-aware
\microbenchmark{} in which each refactoring item is exercised under five-level workloads (XS--XL), with repeated executions for energy
measurement. As we later show in
Section~\ref{subsubsec:result_rq1_microbenchmark}, 29 of the
\microeligiblenumber{} eligible refactoring instances (45.3\%) receive different increase, decrease, or unchanged classifications across XS--XL.
These results demonstrate that relying on a single simple or fixed workload
may miss or even reverse the inferred energy impact of a refactoring.

\noindent\textbf{Limitation 2: Lack of energy measurement on real-world refactoring commits.}
Existing refactoring-energy studies mainly rely on manually constructed refactoring examples or simple example code. For example, \citet{10.1145/2652524.2652538} selected six commonly used Eclipse refactoring types, applied one refactoring type at a selected program location, and then measured the energy consumption of the original and refactored versions. Similarly, \citet{park2014investigation} constructed small original and refactored examples for Fowler refactoring types. These designs are useful for isolating the effect of a refactoring type, but they do not directly represent how refactoring appears in software evolution history.

In real-world development, refactoring is often performed at commit level rather than as a single isolated transformation. A refactoring commit may modify multiple files and contain multiple refactoring instances. Its energy impact may therefore differ from that observed in a manually constructed single-refactoring example. This limitation motivates our construction of the \practicalbenchmark{}, where we study the energy impact of real-world refactoring commits.

\noindent\textbf{Limitation 3: Limited understanding of the causes of refactoring-induced energy differences.} Measuring whether refactoring changes energy consumption is important, but measurement alone does not explain why the change occurs. To make refactoring-energy results actionable, it is necessary to associate observed energy differences with concrete source-level and runtime-level factors. Such explanations can help developers understand which types of refactoring are more energy-sensitive, guide the design of energy-aware refactoring recommendations, and inspire future techniques for detecting or optimizing energy-related regressions. Prior studies on software performance and energy optimization have shown that identifying recurring code patterns and runtime behaviors is useful for building practical detection and optimization techniques~\cite{10.1145/3643991.3644920,10.1145/2568225.2568229,10.1145/2254064.2254075}.

Existing refactoring-energy studies mainly report whether energy consumption changes after refactoring, but provide limited explanation of the causes behind such changes. For example, \citet{10.1145/2652524.2652538} found that refactoring can increase or decrease energy consumption, and further showed that execution time and dynamic execution counts are insufficient to accurately predict the energy impact of applying a refactoring. This result indicates that refactoring-induced energy differences cannot be fully explained by a single simple metric. However, the underlying factors, such as changes in allocation behavior and garbage collection, remain underexplored. This limitation motivates our investigation of the factors that help explain why refactoring leads to energy consumption differences (\rqthree{}).

\noindent\textbf{Limitation 4: Lack of evaluation of existing techniques for detecting refactoring-induced energy regressions.}
Energy measurement is expensive and difficult to apply to every refactoring change during development. If energy-regression refactoring can be identified before full energy measurement, developers could prioritize risky refactoring commits for further validation. This is important because energy regressions may not be visible from functional behavior: a refactored program may preserve the same outputs while consuming more energy during execution. Therefore, beyond measuring and explaining energy differences, it is also important to study whether existing regression-detection techniques can help identify refactoring-induced energy regression.

Existing refactoring-energy studies provide limited evidence on this problem. For example, \citet{10.1145/2652524.2652538} examined whether easily collectible information, execution time and dynamic execution counts, can characterize the energy impact of refactoring, but found that these two signals are insufficient to predict the effect of applying a refactoring. This result suggests that detecting refactoring-induced energy regressions may require richer information than simple runtime or execution-count metrics. However, it remains unclear how well existing regression-prediction techniques can identify refactoring commits that significantly increase energy consumption. This limitation motivates our evaluation of metric-based techniques adapted from existing regression-prediction studies~\cite{10.1145/3643991.3644920} and our further exploration of whether LLMs can help identify refactoring-induced energy regressions (\rqfour{}).

To address these limitations, we conduct an empirical study of the energy impact of software refactoring from two complementary perspectives. First, we construct \emph{\microbenchmark{}}, a workload-aware benchmark of executable Java refactoring examples. It covers \micrototalnumber{} refactoring types and exercises the refactored code under diverse inputs and repeated executions, allowing us to study how individual refactoring types affect energy consumption under controlled but non-trivial workloads. Second, we construct \emph{\practicalbenchmark{}}, a benchmark of real-world GitHub refactoring commits. This benchmark allows us to measure the energy impact of refactoring as it appears in software evolution history, where a refactoring commit may modify multiple files and contain multiple refactoring instances. Across both benchmarks, we measure the energy consumption before and after refactoring, apply paired statistical analysis, and further examine factors that help explain energy differences and techniques that identify energy regression refactoring commits. Specifically, we investigate the following research questions:

\begin{description}[leftmargin=0pt]

\item[RQ1: What is the impact of refactoring on energy consumption?] We measure and compare the energy consumption of original and refactored programs in both \microbenchmark{} and \practicalbenchmark{}, and analyze whether refactoring leads to statistically significant energy differences.

\item[RQ2: Which refactoring types are more likely to influence energy consumption?]  We further analyze energy changes across refactoring types in \microbenchmark{} and examine refactoring co-occurrence patterns in \practicalbenchmark{} to understand which refactoring types or combinations are more associated with energy differences.

\item[RQ3: What factors explain refactoring-induced energy differences?] We investigate source-level and runtime-level factors, such as execution time, allocation behavior, garbage collection, and compiler optimization behavior, to understand why refactoring can lead to energy consumption differences.

\item[RQ4: How effective are existing techniques in identifying refactoring-induced energy regressions?] We evaluate metric-based techniques adapted from existing regression-prediction studies~\cite{10.1145/3643991.3644920} and further explore whether LLMs can identify refactoring commits that significantly increase energy consumption.

\end{description}

\noindent\textbf{Key Findings.}
Our empirical study shows that refactoring is not inherently energy-neutral. Across controlled benchmark (\microbenchmark{}), 51.8\% of refactoring--workload comparisons exhibit statistically significant energy differences, and over one-quarter (27.9\%) change energy consumption by at least 10\%. However, the observed impact is highly workload-dependent: nearly half (45.3\%) of refactorings change their inferred energy outcome across workloads, demonstrating that conclusions drawn from a single workload can be misleading. In contrast, statistically significant energy changes occur less frequently in real-world refactoring commits (7.5\%), although those changes are often substantial, with two-thirds (66.7\%) exceeding 10\%. We further show that while many refactoring types and recurring refactoring combinations are associated with significant energy differences, their effects are not deterministic and depend on execution context. Finally, execution time explains only part of the observed energy variation, and neither metric-based approaches nor LLM-based predictors reliably identify energy regressions. Overall, these findings demonstrate that the energy impact of refactoring is complex, context-dependent, and difficult to predict without direct empirical measurement.

This paper makes the following contributions:

\begin{itemize}
    \item We construct \emph{\microbenchmark{}}, a workload-aware benchmark of executable Java refactoring examples covering \micrototalnumber{} refactoring types. This benchmark enables controlled energy analysis of individual refactoring types under diverse workloads.

    \item We construct \emph{\practicalbenchmark{}}, a benchmark of real-world GitHub refactoring commits. This benchmark enables the study of refactoring energy impact as it appears in software evolution history, where commits may involve multiple files and multiple refactoring operations.

    \item We conduct a systematic empirical study of refactoring energy impact across both controlled examples and real-world commits. Our study measures before- and after-refactoring energy consumption under repeated executions and applies statistical analysis to quantify refactoring-induced energy differences. To support open science, the code and data of our study are publicly available at ~\cite{opensourcelink}.

    \item We evaluate whether refactoring-induced energy regressions can be identified using metric-based techniques adapted from regression-prediction studies, and further explore the capability of LLM for this task.
\end{itemize}
\section{Motivation Example}
\label{sec:examples}

\begin{figure}[ht]
\centering
\noindent
\begin{minipage}[b]{0.6\linewidth}
\centering
\begin{lstlisting}[style=mystyle, frame=single, numbers=left, mathescape, caption={Code before refactoring.}, label=beforepipelinerefactoring]
 public static final class Names {
    public List<String> namesOfProgrammers(List<Person> input) {
-       List<String> names = new ArrayList<>();
-       for (Person p : input) {
-           if ("programmer".equals(p.getJob())) {
-               names.add(p.getName());
-           }
-       }
-       return names;
    }
 }
\end{lstlisting}
\end{minipage}

\begin{minipage}[b]{0.6\linewidth}
\centering
\begin{lstlisting}[style=mystyle, frame=single, numbers=left, mathescape, caption={Code after refactoring.}, label=afterpipelinerefactoring]
 public static final class Names {
    public List<String> namesOfProgrammers(List<Person> input) {
+       return input.stream()
+               .filter(p -> "programmer".equals(p.getJob()))
+               .map(After.Person::getName)
+               .collect(Collectors.toList());
    }
 }
\end{lstlisting}
\end{minipage}
\caption{Example of \textit{Replace Loop with Pipeline} refactoring which increases energy consumption across different workload groups.}
\Description{}
\label{fig:replace_loop_with_pipeline_example}
\end{figure}

Figure~\ref{fig:replace_loop_with_pipeline_example} shows the \textit{Replace Loop with Pipeline} refactoring used as the motivating example from our \microbenchmark{}. The method takes a list of \texttt{Person} objects and returns the names of people whose job is \texttt{"programmer"}. As shown in Listing~\ref{beforepipelinerefactoring}, before refactoring, the method creates an \texttt{ArrayList}, traverses the input list with a \texttt{for} loop, checks whether each person is a programmer, and appends the selected names to the result list. After refactoring, the same computation is expressed with a stream pipeline\footnote{\url{https://docs.oracle.com/javase/8/docs/api/java/util/stream/Stream.html}} that applies \texttt{filter}, \texttt{map}, and \texttt{collect} as shown in Listing~\ref{afterpipelinerefactoring}. The output of the method remains the same, but the execution is no longer the same direct loop-and-append procedure. The refactored version delegates traversal, filtering, mapping, and result construction to the \texttt{stream()} API. This change 
captures the case where a refactoring improves code style or readability, but may also change the runtime behavior of the program.

\begin{table}[ht]
\centering
\caption{Workload input scale and coverage summary for the motivating \textit{Replace Loop with Pipeline} example.}
\label{tab:motivation_replace_loop_pipeline_workload}
\footnotesize
\begin{adjustbox}{width=0.6\textwidth}
\begin{tabular}{@{}ll@{}}
\toprule
\textbf{Aspect} & \textbf{Value} \\
\midrule
Input type & \texttt{List} \\
Input-size levels & XS=100, S=1{,}000, M=10{,}000, L=100{,}000, XL=1{,}000{,}000 \\
Modified-line coverage & 100\% for both versions \\
\bottomrule
\end{tabular}
\end{adjustbox}
\end{table}

To make the energy comparison meaningful, the workload must execute the changed code and run long enough to produce observable energy measurements. Table~\ref{tab:motivation_replace_loop_pipeline_workload} summarizes the workload setup for this example. The input is a \texttt{List}, and the workload suite covers five input-size levels, from XS with 100 elements to XL with 1,000,000 elements~\cite{10.1145/3030207.3030221,10.1145/3168825}. In addition to reporting each size level separately, we also report an \emph{Aggregate} workload group. This group executes the full workload suite across all input-size levels, and therefore represents the overall behavior of the example when the diverse workloads are considered together. The workloads achieve 100\% modified-line coverage for both versions, ensuring that the refactored logic is exercised during measurement.

For each workload group, we first estimate how many repetitions are needed to make the execution last around 10 seconds, so that the energy consumption is large enough to be measured more reliably~\cite{10.1145/3030207.3030221,10.1145/3168825}. As the before and after versions may have different execution speeds, we use a common repetition count for the two versions. Specifically, we choose the larger repetition count between the before and after versions, and then run both versions with that same count. This avoids comparing two versions under different amounts of logical work. We repeat the energy measurement 30 times for the before version and 30 times for the after version, and use these repeated measurements for the distributions and statistical comparisons reported below~\cite{energymeasurementlink,10.1145/3696630.3728707}.

\begin{figure}[ht]
    \centering
    \vspace{-3pt}    \includesvg[width=0.6\textwidth]{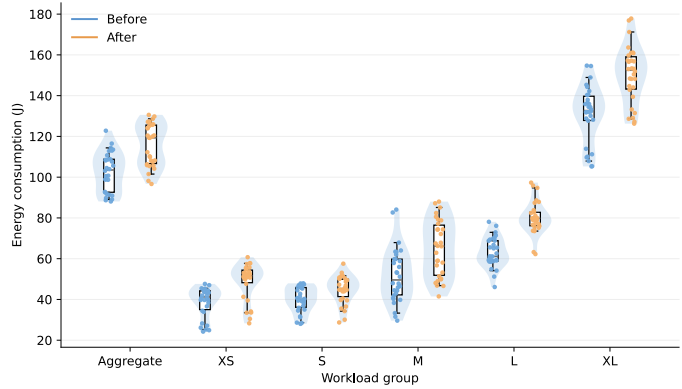}
    \vspace{-6pt}
\caption{Energy consumption distributions before and after the \textit{Replace Loop with Pipeline} refactoring across workload groups.}
\Description{Violin plots comparing before- and after-refactoring energy consumption across the aggregate workload group and five input-size levels.}
\label{fig:example_raw_energy}
\end{figure}

Figure~\ref{fig:example_raw_energy} compares the before- and after-refactoring energy distributions across the Aggregate workload group and the five input-size groups. The after-refactoring distributions are consistently shifted upward, indicating that the stream-based version consumes more energy than the loop-based version under every workload group. The figure also shows that absolute energy consumption generally increases as the workload becomes larger, with XL having the highest energy values for both versions. The Aggregate group summarizes the combined workload suite across XS to XL, so it provides an overall view of the refactoring effect across diverse input scales rather than representing another individual input size. Together, these patterns show that the refactoring increases energy consumption consistently, while the absolute energy cost and distribution shape vary across workload groups.

\begin{figure}[ht]
    \centering
    \vspace{-3pt}    \includesvg[width=0.6\textwidth]{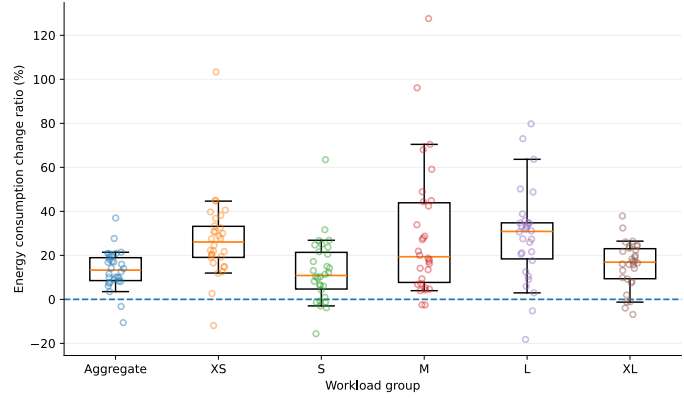}
    \vspace{-6pt}
\caption{Distributions of energy consumption change ratios across workload groups for the motivating \textit{Replace Loop with Pipeline} example. The change ratio is computed as $(E_{\text{after}} - E_{\text{before}})/E_{\text{before}} \times 100$.}
\Description{}
\label{fig:example_energy_change_ratio}
\end{figure}

The change ratios in Figure~\ref{fig:example_energy_change_ratio} show that the increase is not uniform. The Aggregate workload has a mean increase of 12.99\%, while the individual input-size groups range from 11.93\% for S to 26.99\% for L. The XS group is close to L, with a 26.00\% increase. XL has the largest raw energy values, but its relative increase is 14.99\%, smaller than those of L and XS. Thus, a single aggregate value would correctly indicate that the refactoring increases energy, but it would hide how much the measured effect changes with input size. The same refactoring does not correspond to one fixed energy effect across workloads. This motivates the need for workload-aware measurement, since conclusions drawn from one input scale may understate or overstate the impact of a refactoring.

\begin{table}[ht]
\centering
\caption{Energy comparison results for the motivating \textit{Replace Loop with Pipeline} example.}
\label{tab:example_energy_comparison}
\footnotesize
\begin{adjustbox}{width=0.8\textwidth}
\begin{tabular}{@{}lrrrrlcr@{}}
\toprule
\textbf{Workload group} & \textbf{Before (J)} & \textbf{After (J)} & \textbf{$\Delta$ (J)} & \textbf{$\Delta$ (\%)} & \textbf{95\% CI of $\Delta$ (J)} & \textbf{WSR test} & \textbf{Cliff's $\delta$} \\
\midrule
Aggregate & $102.52$ & $115.84$ & $+13.32$ & $+12.99\%$ & $[10.07, 16.58]$ & $W=15,\ p=2.55 \times 10^{-7}$ & $0.60$ \\
XS & $38.81$ & $48.90$ & $+10.09$ & $+26.00\%$ & $[7.90, 12.28]$ & $W=3,\ p=2.35 \times 10^{-6}$ & $0.66$ \\
S & $40.06$ & $44.84$ & $+4.78$ & $+11.93\%$ & $[2.88, 6.68]$ & $W=41,\ p=1.82 \times 10^{-5}$ & $0.40$ \\
M & $51.76$ & $65.12$ & $+13.36$ & $+25.81\%$ & $[9.10, 17.62]$ & $W=5,\ p=1.86 \times 10^{-8}$ & $0.51$ \\
L & $63.07$ & $80.09$ & $+17.02$ & $+26.99\%$ & $[12.77, 21.26]$ & $W=12,\ p=1.30 \times 10^{-7}$ & $0.91$ \\
XL & $130.84$ & $150.45$ & $+19.61$ & $+14.99\%$ & $[14.77, 24.44]$ & $W=12,\ p=1.30 \times 10^{-7}$ & $0.64$ \\
\bottomrule
\end{tabular}
\end{adjustbox}
\begin{flushleft}
\footnotesize Before and after values are means over repeated measurements. $\Delta$ is computed as $E_{\text{after}} - E_{\text{before}}$, and the percentage change is computed as $(E_{\text{after}} - E_{\text{before}})/E_{\text{before}} \times 100$. WSR test denotes the Wilcoxon signed-rank test.
\end{flushleft}
\end{table}

Table~\ref{tab:example_energy_comparison} provides the statistical comparison behind these distributions. The after-refactoring version has a higher mean energy consumption for every workload group. The mean difference is +13.32 J for Aggregate and ranges from +4.78 J to +19.61 J across the individual input-size groups. The 95\% confidence intervals of the mean differences are all positive and the Wilcoxon signed-rank tests are statistically significant for all workload groups. Cliff's $\delta$ ranges from 0.40 to 0.91, indicating non-negligible effects. These results support two conclusions. First, the refactoring is not energy-neutral in this example. Second, both absolute and relative views are needed. XL has the largest absolute mean difference, while L and XS have the largest relative increases. This means that a single workload may not fully characterize the energy impact of the refactoring. To understand refactoring-induced energy changes, measurement needs to consider how the refactored code behaves under different workload scales.

\begin{table}[ht]
\centering
\caption{Selected runtime-profile metrics for the motivating \textit{Replace Loop with Pipeline} example.}
\label{tab:example_runtime_profile}
\footnotesize
\begin{adjustbox}{width=0.88\textwidth}
\begin{tabular}{@{}lrrrrrr@{}}
\toprule
\textbf{Workload group} & \textbf{$\Delta E$ (\%)} & \textbf{$\Delta$ Time (\%)} & \textbf{$\Delta$ Alloc. samples (\%)} & \textbf{$\Delta$ Young GC (\%)} & \textbf{$\Delta$ GC pause (\%)} & \textbf{$\Delta$ CPU-load records (\%)} \\
\midrule
Aggregate & $+12.99$ & $+16.06$ & $+13.29$ & $+4.88$ & $+7.87$ & $+20.00$ \\
XS & $+26.00$ & $+93.57$ & $+86.12$ & $+50.00$ & $+77.14$ & $+114.29$ \\
S & $+11.93$ & $+47.06$ & $+47.88$ & $+21.88$ & $+15.47$ & $+66.67$ \\
M & $+25.81$ & $+38.45$ & $+33.71$ & $+3.90$ & $+4.61$ & $+44.44$ \\
L & $+26.99$ & $+23.55$ & $+19.05$ & $+57.58$ & $+11.42$ & $+30.00$ \\
XL & $+14.99$ & $+26.78$ & $+28.87$ & $+8.70$ & $+36.37$ & $+25.00$ \\
\bottomrule
\end{tabular}
\end{adjustbox}
\begin{flushleft}
\footnotesize All values are computed as $(X_{\text{after}} - X_{\text{before}})/X_{\text{before}} \times 100$. $\Delta E$ compares mean energy consumption across repeated measurements. $\Delta$ Time is the change ratio of execution time. $\Delta$ Alloc. samples compares sampled heap-object allocation events during execution, which reflect objects created by the program. $\Delta$ Young GC compares young-generation garbage-collection events, which mainly reclaim short-lived heap objects. $\Delta$ GC pause compares the total time spent in garbage-collection pauses. $\Delta$ CPU-load records compares the number of recorded CPU-load observations during execution.
\end{flushleft}
\end{table}

The energy results show that the refactoring changes energy consumption, but they do not explain the source of the change. Table~\ref{tab:example_runtime_profile} reports selected runtime-profile metrics for the same example. After refactoring, execution time, allocation samples, young-generation GC events, GC pause time, and CPU-load records all increase. These changes are consistent with the implementation difference in Figure~\ref{fig:replace_loop_with_pipeline_example}: the refactored version replaces a direct loop and explicit list update with stream operations, lambda-based processing, and collector-based result construction. Such changes can introduce additional runtime work even when the method output is preserved.

The runtime-profile metrics also show that the energy increase cannot be explained by a single factor. Under XS, energy increases by 26.00\%, while execution time and allocation samples increase by 93.57\% and 86.12\%, respectively. Under M, the energy increase is 25.81\%, but the young-GC increase is only 3.90\%. Under L, energy increases by a similar amount, 26.99\%, while young GC increases by 57.58\%. These patterns suggest that execution time, allocation behavior, garbage collection, and CPU activity provide complementary signals. Energy measurement reveals that a difference exists, but runtime-profile metrics are needed to help interpret why the difference occurs.

This motivating example highlights several challenges in understanding the energy implications of software refactoring. First, although the refactoring preserves externally observable behavior, it introduces a statistically significant energy change, demonstrating that the energy impact cannot be inferred from functional equivalence alone and must be empirically measured (RQ1). Second, the magnitude and direction of the energy change might vary across workload scales, highlighting the importance of workload-aware evaluation rather than relying on a single execution scenario. Third, this is only one refactoring instance, therefore, a broader analysis is needed to characterize the relationship between refactoring types, refactoring patterns, and energy changes (RQ2). Finally, the observed energy variation is correlated with changes in execution time, memory allocation, garbage collection, and CPU activity, suggesting that runtime characteristics can provide valuable evidence to explain energy differences (RQ3). These observations motivate our empirical study, which systematically investigates refactoring-induced energy changes across controlled benchmarks and real-world commits, identifies source-level and runtime factors associated with these changes, and evaluates whether available signals can support the detection of refactoring-induced energy regressions (RQ4).

\section{Methodology}
\label{sec:methodology}

\subsection{Study Overview}
\label{subsec:study_overview}

\begin{figure}[!htbp]
    \centering
    \includegraphics[width=0.57\textwidth]{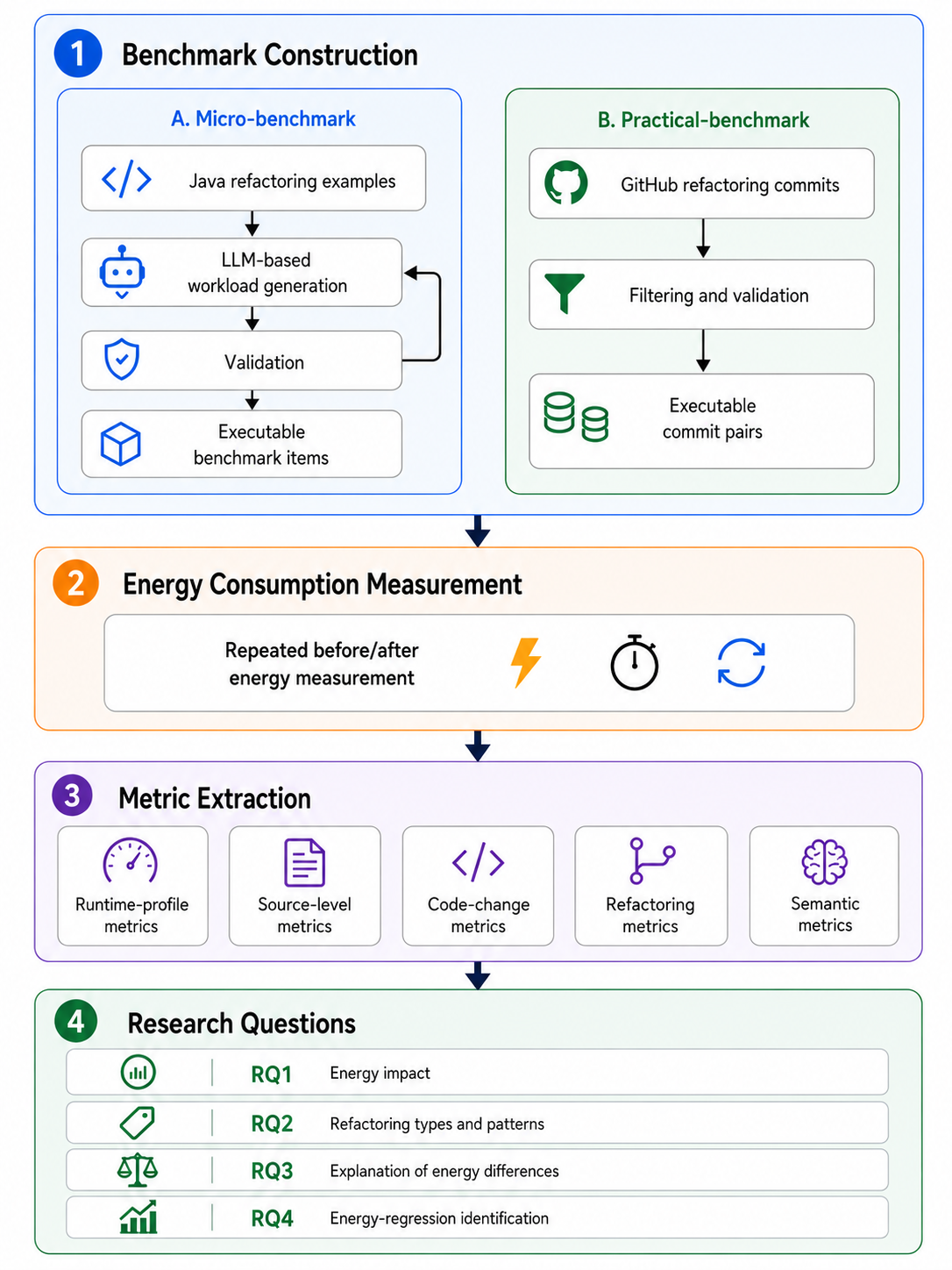}
    \caption{Overview of the study workflow.}
    \label{fig:study_workflow_overview}
    \Description{}
\end{figure}

Figure~\ref{fig:study_workflow_overview} summarizes the overall workflow of our study. The methodology consists of four main steps: benchmark construction, energy consumption measurement, metric extraction, and research-question-oriented analysis. We design the workflow around two complementary benchmarks. The first one, \emph{Micro-benchmark}, provides controlled Java refactoring examples for studying individual refactoring types under workload-aware executions. The second one, \emph{Practical-benchmark}, contains real-world GitHub refactoring commits for studying refactoring energy impact as it appears in software evolution history.

In the benchmark construction step, we prepare executable before/after program pairs for both benchmarks. For \emph{Micro-benchmark}, we start from Java refactoring examples, use an LLM-assisted pipeline to generate workloads, and validate the generated benchmark items through compilation and coverage checks. If a generated workload fails validation, the pipeline regenerates it until a valid executable benchmark item is obtained or the retry limit is reached. For \emph{Practical-benchmark}, we collect candidate GitHub refactoring commits and apply filtering and validation steps to obtain executable commit pairs whose before and after versions can be measured under comparable test workloads. The detailed construction procedures are described in Section~\ref{subsec:benchmark_construction}.

After benchmark construction, we measure energy consumption by repeatedly executing the before and after versions of each benchmark subject. This measurement step records per-run energy and timing information, which is later used to compare the energy behavior before and after refactoring. For \emph{Micro-benchmark}, the measurement uses generated workloads and a common repetition setting for the before and after versions so that the two versions execute the same amount of logical work. For \emph{Practical-benchmark}, the measurement uses project test suites as workloads. The detailed energy measurement strategy is described in Section~\ref{subsec:energy_measurement_procedure}.

We also collect metrics to support explanation and regression-identification analyses. Runtime-profile metrics characterize how the program executes under the benchmark workload, such as execution time, memory allocation activity, garbage collection, and CPU-related observations. In addition, source-level, code-change, refactoring, and semantic metrics are extracted from the program source code and commit diffs to describe the structure and content of the refactoring changes. These metrics provide complementary views for analyzing why refactoring can lead to energy differences and whether such differences can be identified from measurable signals. Section~\ref{subsec:metric_extraction} describes the metrics and their extraction procedures in detail.

Finally, the collected energy measurements and metrics are used to answer the four research questions. RQ1 compares before/after energy consumption to quantify the energy impact of refactoring. RQ2 analyzes which refactoring types and refactoring patterns are more strongly associated with energy differences. RQ3 examines runtime-level and source-level factors that help interpret observed energy changes. RQ4 evaluates whether metric-based and LLM-based techniques can identify refactoring-induced energy regressions. Section~\ref{sec:experimental_design} presents the experimental design for answering these research questions.

\subsection{Benchmark Construction}
\label{subsec:benchmark_construction}

\subsubsection{Micro-benchmark}
\label{subsubsec:micro_benchmark}

The goal of \emph{Micro-benchmark} is to provide a controlled benchmark for measuring the energy impact of individual refactoring types. In real-world commits, multiple refactoring operations may be applied together, making it difficult to isolate the effect of one refactoring type. Micro-benchmark complements such commit-level analysis by providing small executable Java programs in which the before and after versions implement one intended refactoring type and can be exercised under the same generated workloads.

We construct \emph{Micro-benchmark} based on Fowler's refactoring catalog~\cite{onlinerefactoringcatalog}. We use the second edition catalog as the reference list of refactoring types because it provides a comprehensive and updated organization of refactoring operations~\cite{fowler2018refactoring}. However, the online examples in the second edition are mainly written in JavaScript and many of them are illustrative fragments rather than directly executable Java benchmark programs. Since our measurement pipeline targets Java programs, we use the first edition of Fowler's refactoring book as the primary source for Java-based examples~\cite{fowler1999refactoring}. For each refactoring type that appears in both editions, we match the refactoring by name and manually extract the corresponding Java before- and after-refactoring example from the first edition. For refactoring types that do not have Java examples in the first edition, we manually extract the corresponding JavaScript example from the second-edition online catalog and translate it into Java. After collecting these examples, we inspect and validate all resulting program pairs to ensure that the before and after versions are compilable and implement the intended refactoring. In this way, each retained benchmark item contains a Java before/after program pair associated with one refactoring type.

\paragraph{Workload generation.}
The Java before/after program pairs extracted from Fowler's refactoring examples are designed to illustrate refactoring mechanics; they do not provide workloads for repeated energy measurement. However, energy and performance measurements can be sensitive to the workload input-size levels used during execution~\cite{10.1145/2884781.2884869,10.1145/3168825,10.1145/3030207.3030221}. As discussed in Limitation~1, the same refactoring may show different energy effects when the refactored code is executed with different workload input-size levels, as shown in our motivation example in Section~\ref{sec:examples}. To address this issue, we generate diverse workloads for each benchmark item by varying input-size levels.

\begin{figure}[!htbp]
\centering
\includegraphics[width=0.7\textwidth]{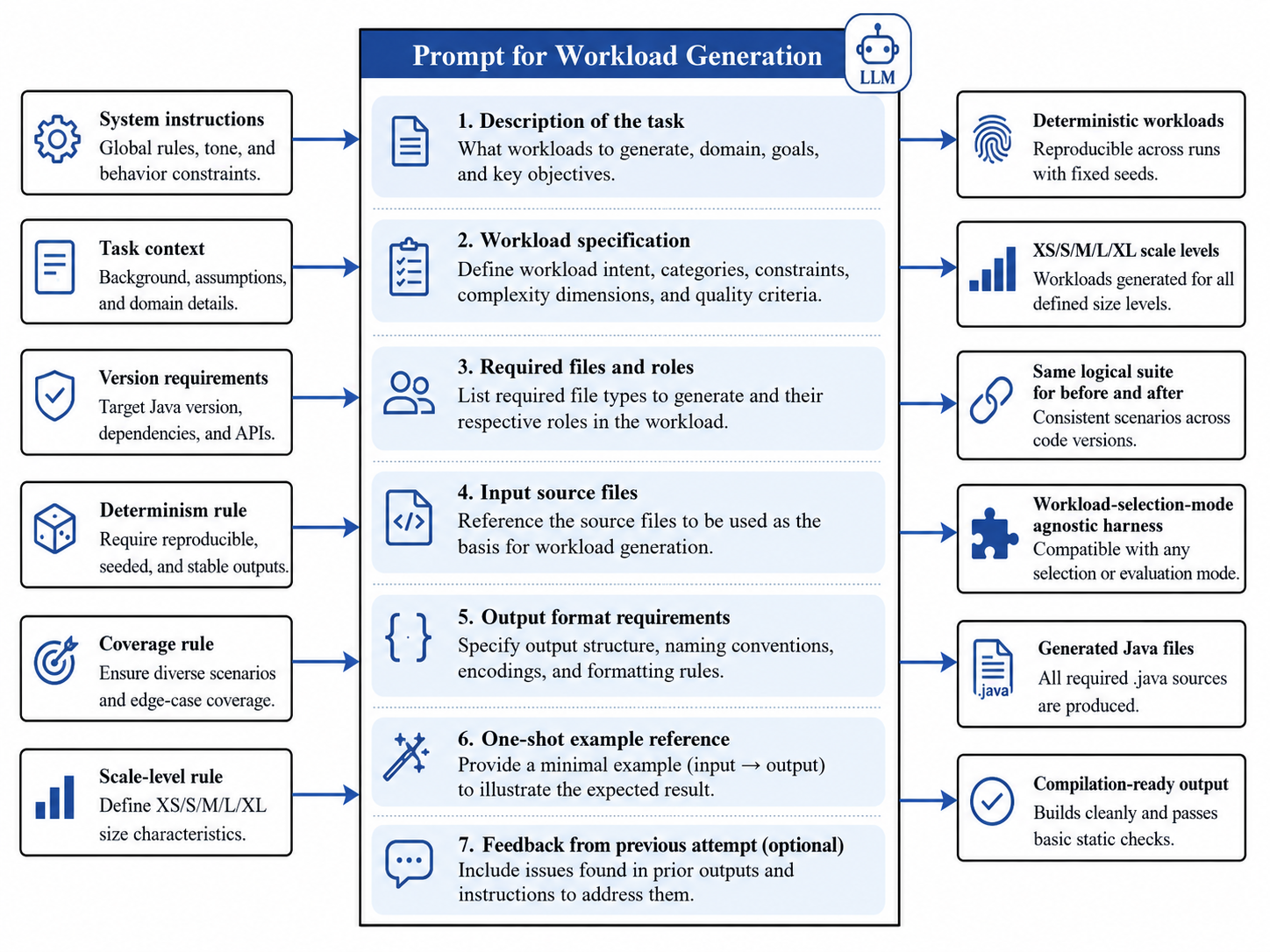}
\caption{Prompt structure for LLM-assisted workload generation in Micro-benchmark.}
\label{fig:workload_generation_prompt}
\Description{}
\end{figure}

We use an LLM with a one-shot example to generate workloads for each refactoring item. Figure~\ref{fig:workload_generation_prompt} summarizes the prompt structure. The prompt is organized into seven main components.

\begin{enumerate}
\item \textit{Description of the task.} This component states that the goal is to generate workloads for one refactoring item. It defines the generation objective: the workloads should exercise the refactored computation and support repeated execution for energy measurement.

\item \textit{Workload specification.} This component gives the main workload-generation requirements. It requires deterministic workloads, a mandatory input-size scale axis with the XS/S/M/L/XL levels, and coverage of the modified logic in both the before and after versions. It also allows additional logic axes only when they directly affect the refactored logic.

\item \textit{Required files and roles.} This component describes the expected output organization and the role of each generated output. Its purpose is to make the generated workload executable by the downstream measurement pipeline, rather than producing only isolated test inputs.

\item \textit{Input source files.} This component provides the before- and after-refactoring Java programs for the current item. The LLM uses these source files to identify the method under measurement, identify relevant input types, and generate workloads that exercise the refactored code.

\item \textit{Output format requirements.} This component specifies the format of the LLM response so that the pipeline can automatically extract and compile the generated output. This reduces manual intervention and enables the same generation procedure to be applied across benchmark items.

\item \textit{One-shot example reference.} This component provides one complete example to illustrate the expected result. The example is used as a reference for workload generation, deterministic input construction, and compatibility with the measurement pipeline, but the generated workloads are adapted to the current refactoring item.

\item \textit{Feedback from previous attempt.} When a generation attempt fails file extraction or compilation, the pipeline includes the corresponding diagnostic information in the next prompt. This feedback allows the LLM to repair the generated output in subsequent attempts.

\end{enumerate}

This prompt design makes the workload generation systematic across refactoring items while still allowing each workload to be adapted to the input types and execution behavior of the corresponding program.

Table~\ref{tab:micro_workload_specification} summarizes the input-size levels used during workload generation. Each workload suite includes one mandatory scale axis with five levels: XS, S, M, L, and XL. We choose these levels as an order-of-magnitude scale rather than as item-specific constants. Prior Java performance and energy studies commonly vary input size to construct performance or energy profiles~\cite{10.1145/2884781.2884869,10.1145/3168825,10.1145/3030207.3030221}. Following this practice, we use five progressively larger levels from $10^2$ to $10^6$ so that each refactoring item can be exercised from small to large inputs under a uniform scale. Depending on the input type of the refactored code, the same scale values may correspond to the number of collection elements, string characters, loop iterations, or map/set entries.

\begin{table}[!htbp]
\centering
\caption{Input-size levels for workload generation in Micro-benchmark.}
\label{tab:micro_workload_specification}
\footnotesize
\begin{adjustbox}{width=0.48\textwidth}
\begin{tabular}{@{}lccccc@{}}
\toprule
\textbf{Scale level} & \textbf{XS} & \textbf{S} & \textbf{M} & \textbf{L} & \textbf{XL} \tabularnewline
\midrule
\textbf{Size value} & $100$ & $1{,}000$ & $10{,}000$ & $100{,}000$ & $1{,}000{,}000$ \tabularnewline
\bottomrule
\end{tabular}
\end{adjustbox}
\end{table}

Besides input size, some refactored code may also be affected by the distribution of input values. Therefore, the prompt allows additional workload characteristics to be varied only when they are relevant to the refactored logic. For example, a workload may vary the proportion of values in a collection that satisfy a condition or the proportion of null values in the input. These additional characteristics are not added by default. We do not predefine these additional characteristics for all benchmark items; instead, we instruct the LLM to identify them from the current before/after program pair when constructing the workloads.

The generated workloads follow several design requirements. They must exercise the modified logic in both the before and after versions, be deterministic and reproducible. They also support both aggregate execution over all size levels and separate execution for individual size levels. The aggregate execution provides an overall view of the refactoring item's energy behavior, while the per-size execution allows us to examine whether the energy effect changes across input-size levels as shown in the motivation example in Section~\ref{sec:examples}.

\paragraph{Validation of generated workloads.}
After each LLM generation attempt, the pipeline applies several validation checks to the generated output. It first checks whether the LLM response contains all required files. It then compiles the generated Java files to check that the before- and after-refactoring programs and the generated workloads are syntactically valid and can be built together. If file extraction or compilation fails, the pipeline feeds the diagnostic information back to the LLM and asks it to regenerate the output, up to a predefined retry limit.

For workloads that compile successfully, we further check whether they execute the refactored code. We use \emph{modified-line coverage} as the validation criterion: the workload must cover the executable lines changed by the refactoring in both the before and after versions. This check helps prevent us from using workloads that compile and run but do not exercise the modified logic.

After these automated checks, we manually inspect the generated workloads as a final quality-assurance step. The inspection focuses on whether the workload inputs are meaningful for the program under measurement, whether the workloads exercise the computation affected by the refactoring, and whether the input-size levels and any additional workload characteristics are consistent with the program's input type and modified logic. If the inspection identifies an obvious mismatch, we ask the LLM to regenerate the workloads with diagnostic information and then rerun the automatic validation checks. Items that cannot pass both the automatic validation checks and the final manual inspection are excluded. Only the remaining items are included as executable \emph{Micro-benchmark} items.

The final \emph{Micro-benchmark} contains executable Java benchmark items covering 68 refactoring types. Each item consists of a before-refactoring program, an after-refactoring program, and generated workloads that support both aggregate execution over all input-size levels and separate execution for individual input-size levels. This benchmark enables controlled before/after energy comparison for individual refactoring types and supports the input-size-aware analysis used in RQ1 and RQ2.

\subsubsection{Practical-benchmark}
\label{subsubsec:practical_benchmark}

The refactoring examples in \emph{Micro-benchmark} are intentionally small and controlled. This design helps isolate individual refactoring types, but it does not capture the complexity of refactoring in real software evolution. In practice, developers may apply multiple refactoring operations in the same commit, and the surrounding context, dependencies, and test workload may all influence the observed energy behavior. To complement the controlled setting of \emph{Micro-benchmark}, we construct \emph{Practical-benchmark}, a benchmark of real-world Java refactoring commits collected from GitHub repositories.

\paragraph{Candidate commit collection.}
We collect candidate refactoring commits from existing refactoring datasets and empirical studies~\cite{9136878,liu2025automated,nyirongo2025empirical,10.1145/3715730}. For each candidate commit, we retrieve its parent commit and construct a before/after pair, where the parent commit is treated as the before-refactoring version and the candidate commit is treated as the after-refactoring version. We exclude merge commits because they have multiple parents and do not provide a clean one-to-one comparison between the before and after versions.

\paragraph{Precondition checking.}
We apply several precondition checks to ensure that each commit pair can support controlled energy measurement. First, we restrict the benchmark to Maven-based Java projects and require the project to contain a root-level \texttt{pom.xml} file. Second, we require both the before and after versions to compile successfully. Third, we exclude commits that modify test files, because changes to tests may change the test oracle used to compare the two versions.

We then check whether the project tests exercise the modified code. We run the project test suite with coverage collection and compute the intersection between covered lines and the executable lines changed by the commit. If the tests do not cover the modified code, the commit is excluded because its energy impact would not be observable under the test workload. Finally, we require the before and after versions to pass the project tests under the same test suite. These checks ensure that the two versions can be built, tested, and measured under comparable conditions.

\paragraph{Manual refactoring purity inspection.}
After the automated precondition checking, we manually inspect the remaining candidate commits to further reduce the risk of including tangled or behavior-changing commits~\cite{10.1145/3524842.3528034,10.1145/2950290.2950305}. This step follows prior refactoring studies, which use manual inspection of commit messages and code diffs to identify pure refactoring changes and remove commits containing non-refactoring edits~\cite{10.1145/3485136,nyirongo2025empirical}. In our study, two annotators inspected 543 candidate commits that passed the automated checks.

For each candidate commit, the annotators inspected the commit message and the code diff between the before and after versions. Following prior studies~\cite{nyirongo2025empirical}, the goal was to check whether the code changes consisted only of behavior-preserving restructuring rather than functional edits. A commit was classified as a \emph{pure refactoring commit} when its changes only involved refactoring operations, such as extracting, moving, or inlining. A commit was excluded when it introduced new functionality, fixed a bug, or changed program logic. To improve consistency, the two annotators first inspected the commits independently and then discussed disagreements until reaching an agreement. The initial annotations differed for 23 commits, accounting for 4.2\% of the inspected candidates. The two annotators then held a discussion meeting to review these inconsistent cases and reached an agreement on all of them. After the manual inspection, we excluded 62 commits that were not pure refactoring commits. The remaining 481 commits form the final \emph{Practical-benchmark}. The manual inspection took approximately 120 person-hours.

After applying the collection, precondition-checking, and manual-inspection steps, \emph{Practical-benchmark} contains real-world refactoring commits whose before and after versions are compilable, whose tests are unchanged, whose modified code is exercised by the test workload, and whose code changes are manually checked as pure refactoring changes. Unlike \emph{Micro-benchmark}, an item in \emph{Practical-benchmark} may contain multiple refactoring operations and may modify multiple files. This is intentional: the benchmark is designed to study the energy impact of refactoring as it appears in real software evolution. The final \emph{Practical-benchmark} contains 481 refactoring commits from 430 GitHub projects and supports the commit-level analyses used in RQ1--RQ4.

\subsection{Energy Measurement Procedure}
\label{subsec:energy_measurement_procedure}

After constructing \emph{Micro-benchmark} and \emph{Practical-benchmark}, we measure the energy consumption of each benchmark item before and after refactoring. We use JoularJX~\cite{noureddine-ie-2022} to measure the energy consumption. JoularJX is a Java-based software power-monitoring tool that reports energy consumption. It supports multiple operating systems and builds on platform-specific power interfaces, such as RAPL on Linux and powermetrics on macOS. Following established energy-measurement guidelines~\cite{energymeasurementlink,10.1145/3696630.3728707}, we adopt several strategies to make the measurements reliable. We run the experiments on a dedicated machine, avoid unrelated workloads, and keep the runtime configuration fixed across measurements. Before collecting measurements, we perform a warm-up run. Between consecutive measurements, we insert a cool-down interval. The before- and after-refactoring versions of each benchmark item are executed under identical workloads and the same number of repetitions, ensuring that observed energy differences reflect refactoring-induced changes rather than variations in executed workloads. Each version is measured repeatedly to mitigate runtime variability and obtain sufficient observations for statistical analysis. In particular, we collect 30 
independent energy measurements for each version, following recommendations that repeated measurements are necessary for reliable energy comparison~\cite{energymeasurementlink}. We also randomize the measurement order of the before and after versions within each pair, so that one version is not systematically measured earlier or later than the other. The measurement process is automated to ensure that repeated executions follow the same procedure.

\subsubsection{Micro-benchmark Energy Measurement}
\label{subsubsec:micro_energy_measurement}

For \emph{Micro-benchmark}, each benchmark item contains a before-refactoring program, an after-refactoring program, and generated workloads. The generated workloads support two execution modes. The first mode is aggregate execution, where the full workload suite across all input-size levels is executed together. The second mode is per-size execution, where each input-size level is executed separately. This design allows us to measure both the overall energy behavior of a refactoring item and its behavior under individual input-size levels.

For each workload group, the measurement pipeline first determines the repetition count required to ensure that a single measurement run is sufficiently long for reliable energy estimation~\cite{10.1145/3030207.3030221,10.1145/3168825}. Since refactoring may alter execution speed, independently selecting repetition counts for the before and after versions could result in different amounts of executed work and introduce measurement bias. To ensure a fair comparison, we determine the repetition count for each version separately and use the maximum value as the common repetition count for both versions. Consequently, the pre- and post-refactoring versions execute identical generated workloads with the same number of repetitions, allowing measured energy differences to be attributed to refactoring-related changes rather than differences in execution volume.

After the repetition count is determined, we repeatedly measure the energy consumption of the before and after versions. For each workload group, we collect 30 measurements for the before version and 30 measurements for the after version. These repeated measurements form the paired before/after samples used later for statistical comparison.

\subsubsection{Practical-benchmark Energy Measurement}
\label{subsubsec:practical_energy_measurement}

For \emph{Practical-benchmark}, each benchmark item is a real-world refactoring commit. The parent commit is measured as the before-refactoring version, and the refactoring commit is measured as the after-refactoring version. Unlike \emph{Micro-benchmark}, where workloads are generated, \emph{Practical-benchmark} uses the project's existing test suite as the measurement workload, following prior empirical studies that use tests as practical workloads for measuring software runtime or energy behavior~\cite{10.1145/3485136,10.1145/2652524.2652538,hindle2015green}.

For each refactoring commit, the measurement pipeline first prepares the before and after versions, including checking out and building each version. These preparation steps are completed before energy measurement starts. During measurement, we measure only the energy consumption of test-suite execution. We use the same test command for the before and after versions so that the measured energy difference is not caused by different test workloads. The resulting measurements are used in the later experimental analysis. For \emph{Practical-benchmark}, these measurements allow us to study refactoring energy impact at commit level, where a commit may contain multiple refactoring operations and modify multiple files.

\subsection{Metric Extraction}
\label{subsec:metric_extraction}

Energy measurements indicate whether the after-refactoring version consumes more or less energy than the before-refactoring version, but they do not by themselves explain why the difference occurs. To support the explanatory analysis in RQ3 and the regression-identification analysis in RQ4, we extract two groups of complementary metrics. The first group captures runtime behavior, such as execution time, memory allocation, and garbage collection. The second group includes source-level, code-change, coverage, refactoring, and semantic metrics. Together, these metrics describe the runtime behavior, program structure, the amount and distribution of code modifications, how the modified code is exercised, which refactoring operations are applied, and the semantic content of the change. We extract different metrics according to the roles of the two benchmarks. For
\microbenchmark{}, we collect runtime-profile metrics for the controlled
analysis in RQ3. For \practicalbenchmark{}, we collect execution time
together with the other types of metrics for RQ3 and RQ4. Table~\ref{tab:extracted_metrics} summarizes the metric families used in our study.

\begin{table*}[!htbp]
\centering
\caption{Overview of extracted metric families.}
\label{tab:extracted_metrics}
\begin{adjustbox}{width=0.78\textwidth}
\begin{tabular}{l|l|l}
\hline
\textbf{Category} & \textbf{Metrics} & \textbf{Description} \\
\hline\hline
\multirow{4}{*}{Runtime-profile metrics~\cite{10.1145/3669940.3707217,10.1145/1167473.1167488}}
& Execution-time & Time spent executing the workload. \\
& Allocation and memory metrics & Object-allocation activity and memory-usage observations. \\
& Garbage-collection metrics & Garbage-collection events and pause time. \\
& Other runtime metrics & Other runtime-profile observations. \\
\hline
\multirow{3}{*}{Source-level metrics~\cite{9197704,10.1145/3643991.3644920,10.1145/3377811.3380351}}
& CK metrics & Class-level code metrics obtained by static analysis. \\
& Complexity metrics & Cyclomatic complexity and cognitive complexity. \\
& Performance-related static metrics & Static code metrics related to performance-sensitive constructs. \\
\hline
\multirow{8}{*}{Code change metrics~\cite{hindle2015green,10.1145/3377811.3380351}}
& Code churn & Sum of lines added and deleted. \\
& Entropy & Distribution of modified code across files. \\
& Lines added & Total number of lines added in a commit. \\
& Lines deleted & Total number of lines deleted in a commit. \\
& Changed hunks & Number of changed code hunks. \\
& Changed dirs & Number of modified directories. \\
& Changed files & Number of modified files. \\
& Changed subsystems & Number of modified subsystems. \\
\hline
\multirow{1}{*}{Coverage metrics~\cite{10.1145/3485136,bree2021automated}}
& Line coverage & Line coverage ratio for modified files and lines. \\
\hline
\multirow{3}{*}{Refactoring metrics~\cite{10.1145/3485136,10.1145/2652524.2652538,6976079}}
& Applied refactoring types and counts & Set of detected refactoring types and the count of each type. \\
& Refactoring operations total & Total number of detected refactoring operations. \\
& Distinct refactoring types & Number of distinct refactoring types. \\
\hline
\multirow{1}{*}{Semantic metrics~\cite{10174199}}
& Code diff & Code diff for a commit. \\
\hline
\end{tabular}
\end{adjustbox}
\end{table*}

\subsubsection{Runtime-profile Metrics}
\label{subsubsec:runtime_profile_metrics}

Runtime-profile metrics describe how the before and after versions behave at runtime~\cite{10.1145/3669940.3707217,10.1145/1167473.1167488}. We extract these metrics because refactoring may preserve program output while changing execution behavior, such as running time, memory allocation, garbage collection, or other JVM runtime behavior. These runtime changes can help explain why a refactoring leads to an energy increase or decrease.

We extract runtime-profile metrics in separate profiling runs rather than during energy measurement. This separation is important because profiling can introduce additional overhead and may interfere with energy measurement. Therefore, energy measurement and runtime profiling are conducted as separate steps. The profiling step uses the same versions and workload settings as the energy measurement step, but its purpose is to collect runtime-behavior evidence rather than energy values. We focus on representative runtime-profile metrics, including execution-time, heap-object allocation event, young-generation garbage-collection event, total time spent in garbage-collection pauses, and the number of recorded CPU-load observations. The concrete profiling tools are described later in the experimental setup.

For each runtime-profile metric, we compare the after-refactoring value with the before-refactoring value. These runtime-profile metrics are not used as replacements for energy measurement. Instead, they provide complementary evidence for interpreting the measured energy differences. For example, if a refactoring increases both energy consumption and allocation activity, this may suggest that additional object creation contributes to the observed energy increase. Similarly, changes in garbage collection or other runtime behavior can help explain energy differences that are not fully reflected by execution time alone.

\subsubsection{Source-level, Change, Coverage, Refactoring, and Semantic Metrics}
\label{subsubsec:source_change_refactoring_semantic_metrics}

In addition to runtime-profile metrics, we extract metrics that describe the code structure, code change, test coverage, applied refactorings, and semantic content of the change. These metrics are especially important for \emph{Practical-benchmark}, where each item is a real-world commit and may contain multiple refactoring operations. Previous work on commit-level performance regression prediction and performance-aware testing has shown that combining signals from multiple perspectives can improve risk identification compared with using a single metric family alone~\cite{10.1145/3643991.3644920,10.1145/3377811.3380351,9197704}.

\paragraph{Source-level metrics.}
Source-level metrics characterize the structure of the code before and after refactoring. We extract CK metrics, code complexity, and performance-related static metrics~\cite{9197704,10.1145/3643991.3644920,10.1145/3377811.3380351}. CK metrics capture class-level structure, such as coupling and cohesion. Complexity metrics approximate control-flow complexity. Performance-related static metrics capture static constructs that are commonly associated with performance sensitivity~\cite{10.1145/3643991.3644920}. These metrics help describe whether a refactoring changes structural properties that may affect runtime behavior.

\paragraph{Code change metrics.}
Code change metrics describe how much code is changed and how the change is distributed across the project. We extract metrics such as code churn, change entropy, lines added, lines deleted, changed hunks, and changed files~\cite{hindle2015green,10.1145/3377811.3380351}. These metrics are widely used in defect and regression prediction because larger changes and changes scattered across multiple files or modules can be harder to reason about and may be more likely to introduce unintended effects. In our study, these metrics help characterize whether refactoring-related energy differences are associated with the amount and distribution of code modifications.

\paragraph{Coverage metrics.}
Coverage metrics describe whether the changed code is exercised by the measurement workload. We compute line coverage for modified files and modified lines using JaCoCo~\cite{jacocolink}. These metrics are useful because energy differences can only be observed when the workload executes the relevant changed code. For \emph{Practical-benchmark}, coverage metrics indicate what fraction of the modified code is executed by the project test suite, which helps interpret whether the measured energy behavior is likely to reflect the refactoring change.

\paragraph{Refactoring metrics.}
Refactoring metrics explicitly describe which refactoring operations appear in a commit. For each commit in \emph{Practical-benchmark}, we extract the set of applied refactoring types together with the count of each type~\cite{10.1145/3485136,10.1145/2652524.2652538,6976079}. From these counts, we derive the total number of detected refactoring operations and the number of distinct refactoring types. These metrics provide refactoring-specific information that is not captured by generic source-level or change metrics. They are used to analyze whether certain refactoring types or combinations of refactoring types are more associated with energy increases or decreases.

\paragraph{Semantic metrics.}
Count-based metrics cannot fully capture what a code change does. To capture the semantic content of the refactoring edits, we extract the code diff for each refactoring commit~\cite{10174199}. This provides a summary of the transformation content beyond surface statistics, helping distinguish commits that have similar churn or refactoring counts but perform different kinds of changes.

Overall, the extracted metrics provide complementary views of each refactoring change. Runtime-profile metrics describe how the program behaves at runtime, source-level metrics describe the structure of the code, code change metrics describe the amount and distribution of code modifications, coverage metrics describe whether the changed code is executed, refactoring metrics describe the applied refactoring operations, and semantic metrics describe the content of the transformation. This multi-view design supports the explanatory analysis in RQ3 and the regression-identification analysis in RQ4.
\section{Experiments}
\label{sec:experimental_design}

\subsection{Experimental Setup}
\label{subsec:experimental_setup}

All experiments are conducted on a dedicated Mac mini server running macOS with an Apple M4 Pro processor and 24 GB of memory. We use JDK 25 for all benchmark execution. The same machine, operating system, Java runtime, workload, repetition setting, and measurement configuration are used for the before and after versions of each benchmark item. For \practicalbenchmark{}, repository checkout and project build are completed before energy measurement starts; only workload execution is measured.

Runtime-profile metrics are collected separately from energy measurements. This separation avoids introducing profiling overhead into the energy measurements. For runtime profiling, we use JMH together with Java Flight Recorder. The profiling runs use the same machine, JDK version, before/after versions, and workload settings as the corresponding energy measurements, but their outputs are used only to extract runtime-profile metrics for RQ3.

In addition to runtime-profile metrics, we extract the source-level, code change, coverage, refactoring, and semantic metrics described in Section~\ref{subsubsec:source_change_refactoring_semantic_metrics}. For source-level metrics, we use the CK tool~\cite{cklink} to extract CK metrics and complexity metrics, and we use the tools from prior work to extract performance-related static metrics~\cite{9197704,10.1145/3643991.3644920,10.1145/3377811.3380351}. Code change and semantic metrics are extracted from the GitHub REST API for each refactoring commit~\cite{githubapilink}. Coverage metrics are extracted from JaCoCo~\cite{jacocolink} coverage reports. Refactoring metrics are extracted using RefactoringMiner~\cite{refactoringminerlink}, a widely used tool for detecting refactoring operations.

For each benchmark item, the pipeline stores the raw energy-measurement outputs, runtime-profiling outputs, and extracted metric data. The energy-measurement outputs are used to compute before/after energy differences and statistical results. The runtime-profiling outputs are parsed to extract runtime-profile metrics. The source-level, code change, coverage, refactoring, and semantic metric outputs are used in the RQ3 analysis and in the RQ4 regression-identification experiments. The RQ-specific statistical tests, association analyses, and prediction settings are described in the following subsections.

\subsection{RQ1: Overall Energy Impact of Refactoring}
\label{subsec:experiment_rq1}

RQ1 investigates whether refactoring changes energy consumption. We answer this question by comparing the repeated energy measurements of the before-refactoring and after-refactoring versions in both \microbenchmark{} and \practicalbenchmark{}. For each comparison, we use the paired before/after samples collected under the same workload and repetition setting, as described in Section~\ref{subsec:energy_measurement_procedure}.

Let $E^{before}_{i}$ and $E^{after}_{i}$ denote the energy consumption of the before and after versions in the $i$-th paired measurement. We compute the paired energy difference as:
\begin{equation}
\Delta E_i = E^{after}_{i} - E^{before}_{i}.
\end{equation}
We also compute the relative energy change ratio as:
\begin{equation}
\Delta E(\%) = \frac{\overline{E}^{after} - \overline{E}^{before}}{\overline{E}^{before}} \times 100,
\end{equation}
where $\overline{E}^{before}$ and $\overline{E}^{after}$ are the mean energy consumption values across repeated measurements.

To determine whether the before/after energy difference is statistically significant, we apply the paired Wilcoxon signed-rank test~\cite{rosner2006wilcoxon}. We use this non-parametric test because the measurements are paired by construction and energy measurements may not follow a normal distribution. We use $\alpha = 0.05$ as the significance threshold. In addition to $p$-values, we report Cliff's delta~\cite{macbeth2011cliff} as an effect-size measure. Reporting effect size helps distinguish statistically significant differences from changes with limited practical magnitude.

For each comparison, we classify the measured energy impact into three categories. If the Wilcoxon signed-rank test is statistically significant and the mean paired difference is positive, the comparison is classified as an energy increase. If the test is statistically significant and the mean paired difference is negative, the comparison is classified as an energy decrease. Otherwise, the comparison is classified as unchanged. This classification summarizes the direction of the measured effect, while the absolute energy difference, relative change ratio, and effect size report its magnitude. The unit of comparison differs between the two benchmarks. For \microbenchmark{}, we perform the analysis for the aggregate workload group and for each individual input-size group, which allows us to examine both overall and input-size-specific energy changes. For \practicalbenchmark{}, we perform the analysis at commit level using the project test suite as the workload. The resulting energy-impact categories, effect sizes, and change ratios are used in Section~\ref{subsec:experiment_rq1} to report the overall energy impact of refactoring across the two benchmarks.

\subsection{RQ2: Refactoring Types and Refactoring Patterns}
\label{subsec:experiment_rq2}

RQ2 investigates whether the energy impact of refactoring is associated with particular refactoring types or refactoring patterns. While RQ1 focuses on whether refactoring changes energy consumption overall, RQ2 further examines how these changes are distributed across refactoring types in \microbenchmark{} and across refactoring co-occurrence patterns in \practicalbenchmark{}.

For \microbenchmark{}, each benchmark item is associated with one intended refactoring type. We therefore group the before/after comparisons by refactoring type. For each type, we count how many workload groups are classified as increase, decrease, or unchanged according to the RQ1 procedure. We also summarize the distribution of relative energy change ratios for each refactoring type. This analysis allows us to examine whether some refactoring types tend to be more energy-sensitive than others, and whether their effects are consistent or vary across workload groups.

For \practicalbenchmark{}, the analysis is different because a real-world refactoring commit may contain multiple refactoring operations. We use the refactoring metrics extracted by RefactoringMiner to characterize the refactoring operations in each commit, including the detected refactoring types and their counts. We then analyze whether specific refactoring co-occurrence patterns are associated with statistically significant energy increases or decreases. A co-occurrence pattern is defined as a set of refactoring types that appear together in the same commit. We consider refactoring patterns of size two and three, corresponding to pairs and triples of refactoring types. For each pattern $S$, we compute its support as the number of commits containing all refactoring types in $S$.

For each target outcome \(y\), where \(y\) denotes either a statistically
significant energy increase or decrease, we compute the base rate, confidence,
and lift of each pattern following standard association-rule
analysis~\cite{agrawal1993mining,huang2000fast}. Let \(Y\) denote the
energy-impact classification of a commit. The base rate \(P(Y=y)\) is the
proportion of analyzed commits exhibiting outcome \(y\). The confidence of a
pattern \(S\) for outcome \(y\) is the probability that a commit exhibits
\(y\), given that it contains all refactoring types in \(S\):
\begin{equation}
\mathrm{conf}(S,y) = P(Y=y \mid S).
\end{equation}
The lift compares this confidence with the base rate of the target outcome:
\begin{equation}
\mathrm{lift}(S,y)
= \frac{\mathrm{conf}(S,y)}{P(Y=y)}.
\end{equation}
A lift greater than 1 indicates that commits containing \(S\) are more likely
to exhibit outcome \(y\) than an average commit.

In addition to support, confidence, and lift, we report the odds ratio and
the $\phi$ coefficient as effect-size measures for the association between
the occurrence of \(S\) and the outcome \(Y=y\)~\cite{agresti2013categorical}.
We assess each association using Fisher's exact
test~\cite{agresti2013categorical} and apply the Benjamini--Hochberg
false-discovery-rate correction for multiple testing~\cite{benjamini1995controlling}.
The correction is applied separately for each pattern size and target outcome.
Because statistically significant energy increases and decreases are limited
in number, we interpret the results as associative rather than causal evidence.

\subsection{RQ3: Factors Associated with Energy Differences}
\label{subsec:experiment_rq3}

RQ3 investigates which runtime-level and code-level factors are associated with the energy differences observed after refactoring. While RQ1 measures whether energy consumption changes and RQ2 analyzes how the changes are distributed across refactoring types and patterns, RQ3 further examines factors that help interpret these differences. We use the energy change ratios computed in RQ1 together with the runtime-profile, source-level, code change, coverage, refactoring, and semantic metrics described in Section~\ref{subsec:metric_extraction}. These metric families follow prior studies showing that runtime behavior, source-level properties, change characteristics, coverage, and semantic representations can provide useful signals for performance or regression analysis~\cite{10.1145/1167473.1167488,10.1145/3669940.3707217,9197704,10.1145/3377811.3380351,10.1145/3643991.3644920}.

For runtime-profile metrics, we compare the before-refactoring and after-refactoring values of each metric under the same workload setting. Let $X^{before}$ and $X^{after}$ denote the values of a runtime-profile metric before and after refactoring. We compute the relative change ratio of the metric as:
\begin{equation}
\Delta X(\%) = \frac{\overline{X}^{after} - \overline{X}^{before}}{\overline{X}^{before}} \times 100,
\end{equation}
where $\overline{X}^{before}$ and $\overline{X}^{after}$ are the mean metric values across profiling runs. When the before-refactoring value is zero, we use the absolute difference instead of the relative change ratio to avoid undefined ratios. These metric changes allow us to examine whether energy changes are accompanied by changes in execution time, allocation behavior, garbage collection, or other runtime observations. Runtime-profile analysis is particularly relevant for Java programs because JVM execution behavior, memory allocation, garbage collection, and runtime optimization can affect measured performance and energy behavior~\cite{10.1145/1167473.1167488,10.1145/3669940.3707217}.

We first analyze the association between energy change and runtime-profile changes in \microbenchmark{}. For each benchmark item and workload setting, we pair the energy change ratio with the corresponding runtime-profile change ratios. We then compute the Spearman correlation coefficient between the energy change and each runtime-profile metric change~\cite{hauke2011comparison}. We also analyze the association between energy changes and non-runtime metrics in \practicalbenchmark{}. For source-level metrics, we compare the before- and after-refactoring metric values and compute their changes. For code change, coverage, refactoring, and semantic metrics, we use the extracted commit-level values described in Section~\ref{subsubsec:source_change_refactoring_semantic_metrics}. We examine how these metrics differ across the energy-increase, energy-decrease, and unchanged groups defined in RQ1. This analysis follows prior work showing that static code metrics, change metrics, test-related metrics, and learned representations can help characterize performance regressions and related software-quality risks~\cite{9197704,10.1145/3377811.3380351,10.1145/3643991.3644920}.

In addition, we conduct a JVM execution-mode sensitivity analysis for \microbenchmark{}. Since Java programs can be affected by just-in-time compilation, comparing results under different execution modes can help assess whether the observed refactoring energy impact is sensitive to JVM optimization behavior~\cite{10.1145/1167473.1167488,10.1145/3669940.3707217}. We compare the default JVM execution mode with a JIT-disabled execution mode using \texttt{-Xint}. This analysis is not used as the primary energy result; instead, it examines whether the measured refactoring energy impact changes when JIT compilation is disabled. For each refactoring item, we compare the before energy, after energy, and refactoring energy change ratio under the two execution modes. We also rank refactoring types by the difference in energy change ratio between the two modes to identify refactoring types whose estimated energy impact is more sensitive to JIT behavior.

\subsection{RQ4: Refactoring-induced Energy Regression Identification}
\label{subsec:experiment_rq4}

RQ4 investigates whether refactoring-induced energy regressions can be identified using commit-level metrics. While RQ1--RQ3 measure and interpret energy differences, RQ4 treats energy regression identification as a binary classification problem on \practicalbenchmark{}. We focus on \practicalbenchmark{} because it represents real-world refactoring commits, where developers may need to decide whether a refactoring commit should be prioritized for energy validation.

\subsubsection{Ground-truth Labels}
\label{subsubsec:rq4_ground_truth}

We derive the ground-truth labels from the RQ1 statistical analysis. A refactoring commit is labeled as \emph{Reg} if its after-refactoring version shows a statistically significant energy increase compared with its before-refactoring version. Specifically, a commit is labeled as \emph{Reg} when the paired Wilcoxon signed-rank test~\cite{rosner2006wilcoxon} is significant at $\alpha = 0.05$ and the mean paired energy difference is positive. Commits that do not satisfy this condition are labeled as \emph{Not Reg}. Therefore, the \emph{Not Reg} class includes commits with no statistically significant energy change and commits with statistically significant energy decreases. This labeling follows the goal of RQ4: identifying refactoring commits that may introduce energy regressions. We treat \textsc{Reg} as the positive class and report precision, recall, and F1 for this target class.

\subsubsection{Metric-based Prediction Models}
\label{subsubsec:rq4_metric_based_models}

We first evaluate metric-based prediction models from prior performance regression studies~\cite{10.1145/3377811.3380351,10.1145/3643991.3644920}: Logistic Regression, Random Forest, and XGBoost. Logistic Regression provides a linear baseline~\cite{lavalley2008logistic}, Random Forest captures non-linear feature interactions through an ensemble of decision trees~\cite{breiman2001random}, and XGBoost provides a gradient-boosted tree model that is widely used for tabular prediction tasks~\cite{chen2016xgboost}. Each refactoring commit is represented using the extracted metric families described in Section~\ref{subsec:metric_extraction}, including source-level, code change, coverage, refactoring, and semantic metrics, where semantic metric is represented by a 1,024-dimensional code-diff embedding generated by calling the OpenAI text-embedding-3-large model through the OpenAI API. These metrics provide different views of the refactoring commit, such as code structure, change size and distribution, test coverage, detected refactoring operations, and the semantic content of the diff. To reduce the influence of different feature scales, numeric features are standardized before training. We use ten-fold cross-validation and model-specific class weighting to address class imbalance~\cite{10.1145/3377811.3380351,10.1145/3643991.3644920}. All metric-based models are trained and evaluated using the same data split, so their results are comparable. The models are trained on the extracted metrics and evaluated on their ability to identify Reg commits. We use the metric-based models as baselines to assess whether existing commit-level signals are sufficient for identifying refactoring-induced energy regressions

\subsubsection{LLM-based Predictor}
\label{subsubsec:rq4_llm_based_predictor}

In addition to metric-based models, we evaluate an LLM-based predictor. The motivation is that LLMs may capture relationships among heterogeneous signals that are difficult to model with standard tabular classifiers. For each refactoring commit, we construct a structured prompt containing the available metric families and ask the model to classify the commit as \emph{Reg} or \emph{Not Reg}. The prompt does not include the measured energy values or the ground-truth label. We use \texttt{gpt-5-mini} through the OpenAI API~\cite{openaiapilink} as the LLM-based classifier.

We further conduct an ablation study to examine the contribution of different metric families. We organize the prompt information into five groups: static code metrics, code change metrics, test-related metrics, refactoring metrics, and semantic metrics. We first evaluate the model with all metric groups included. Then, we remove one metric group at a time and re-run the classifier. This leave-one-group-out design allows us to observe whether removing a particular metric family reduces the model's ability to identify energy regressions. The LLM-based predictor is not fine-tuned on our dataset. Instead, it is evaluated as a prompt-based classifier. This design reflects a lightweight use case where developers or researchers may query an LLM with structured information about a refactoring commit without training a project-specific prediction model.

\subsubsection{Evaluation Metrics}
\label{subsubsec:rq4_evaluation_metrics}

We evaluate the classifiers using precision, recall, and F1 score, which are standard metrics for binary classification~\cite{sokolova2009systematic}. Since the goal is to identify energy-regression commits, we treat \emph{Reg} as the positive class. Precision measures how many predicted regressions are true regressions, while recall measures how many true regressions are successfully identified. F1 summarizes the trade-off between precision and recall. For both approaches, we report precision, recall, and F1 for the \textsc{Reg} class because the primary purpose is to identify energy regressions rather than recognize the majority \textsc{Not Reg} class.
\section{Results}
\label{sec:result}

This section presents the empirical results organized by the four research questions. We first report the overall energy impact of refactoring (RQ1), then examine which refactoring types and patterns are associated with energy differences (RQ2). We next analyze runtime-level and code-level factors that help explain the observed differences (RQ3), and finally evaluate whether existing techniques can identify refactoring-induced energy regressions (RQ4).

\subsection{RQ1: What is the impact of refactoring on energy consumption?}
\label{subsec:experiment_rq1}

\subsubsection{Energy Impact across Workload Groups in \microbenchmark{}}
\label{subsubsec:result_rq1_microbenchmark}

\begin{figure}[t]
  \centering
  \includegraphics[width=0.5\linewidth]
  {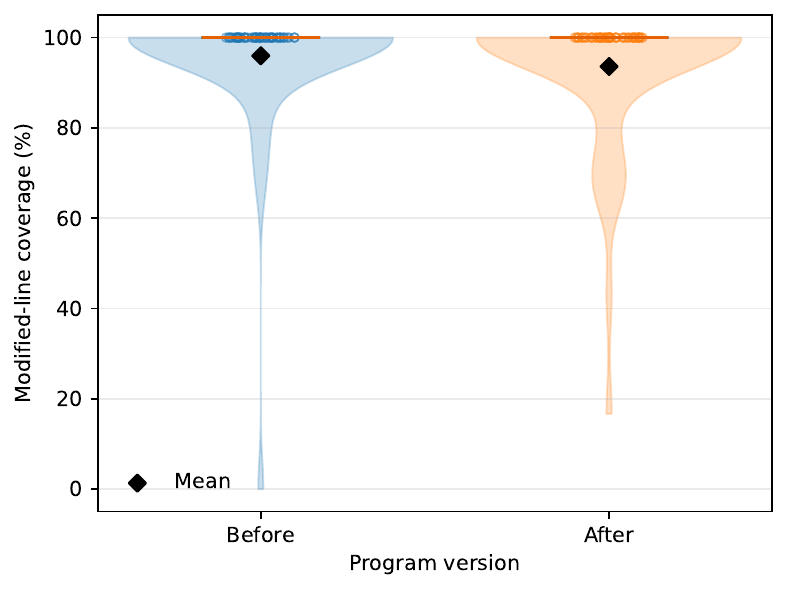}
  \caption{Distributions of modified-line coverage before and after
  refactoring.}
  \label{fig:rq1_micro_coverage_distribution}
\end{figure}

\paragraph{Modified-line coverage.}
To ensure that the measured energy differences are attributable to the refactored code rather than unexecuted changes, we first examine \emph{modified-line coverage}, defined as the proportion of lines modified by a refactoring that are executed by the benchmark workload. High modified-line coverage increases confidence that the measured energy consumption reflects the behavior of the refactored code instead of unaffected portions of the program.
The analysis begins with 71 candidate items. 
An item is included in RQ1 only if it has complete and valid energy measurements for the Aggregate workload and all five input-size workloads. Under this criterion, \microeligiblenumber{} items (90.1\%) are retained, while seven are excluded because the energy-measurement stage could not be completed. The resulting analysis comprises 384 item--workload comparisons (\microeligiblenumber{} items $\times$ six workload groups) and 11,520 matched before/after measurement pairs.
Figure~\ref{fig:rq1_micro_coverage_distribution} summarizes the modified-line coverage of the eligible items. Coverage is highly concentrated at 100\% for both the pre- and post-refactoring versions. Consequently, the median and interquartile range coincide at the upper bound, causing the overlaid boxplots to collapse into nearly indistinguishable horizontal lines. Mean coverage is 95.97\% before refactoring and 93.59\% after refactoring, while the median remains 100\% for both versions. These results indicate that the benchmark workloads exercise nearly all refactored code, reducing the likelihood that the observed energy differences are attributable to insufficient coverage.

\paragraph{Frequency and direction of energy changes.}
Table~\ref{tab:rq1_micro_workload_summary} summarizes how frequently each
energy-impact classification occurs under each workload. A comparison is
statistically significant when its paired Wilcoxon signed-rank test has
$p<0.05$. Significant comparisons are classified as increases or decreases
using the sign of the mean paired difference; therefore, the Significant
column is the sum of Increase and Decrease. Unchanged denotes the absence of a
statistically significant difference rather than an observed difference of
exactly zero. Each workload group contains 64 items, and
all percentages use this number as the denominator. Across the 384 item--workload comparisons, 199 (51.8\%) exhibit a significant
energy change, while 185 (48.2\%) are unchanged. The significant comparisons
comprise 113 increases (29.4\%) and 86 decreases (22.4\%). The proportion of
significant comparisons ranges from 46.9\% for Aggregate to 54.7\% for S and
L, and every workload contains both significant increases and significant
decreases. Among the significant comparisons, no workload contains
significantly more increases than decreases, or vice versa, according to
two-sided exact binomial tests with Holm correction
($p_{\mathrm{Holm}}\geq0.481$).

\begin{table*}[t]
  \centering
  \caption{Energy-change classifications by workload group.}
  \label{tab:rq1_micro_workload_summary}
  \resizebox{0.68\textwidth}{!}{%
    \begin{tabular}{@{}lrrrr@{}}
      \toprule
      \textbf{Workload} &
      \textbf{Unchanged, $n$ (\%)} &
      \textbf{Significant, $n$ (\%)} &
      \textbf{Increase, $n$ (\%)} &
      \textbf{Decrease, $n$ (\%)} \\
      \midrule
      Aggregate & 34 (53.1) & 30 (46.9) & 16 (25.0) & 14 (21.9) \\
      XS        & 32 (50.0) & 32 (50.0) & 18 (28.1) & 14 (21.9) \\
      S         & 29 (45.3) & 35 (54.7) & 18 (28.1) & 17 (26.6) \\
      M         & 31 (48.4) & 33 (51.6) & 22 (34.4) & 11 (17.2) \\
      L         & 29 (45.3) & 35 (54.7) & 22 (34.4) & 13 (20.3) \\
      XL        & 30 (46.9) & 34 (53.1) & 17 (26.6) & 17 (26.6) \\
      \bottomrule
    \end{tabular}%
  }
\end{table*}

\begin{findingbox}
The benchmark workloads achieve consistently high modified-line coverage, indicating that the observed energy differences are attributable to the refactored code rather than unexecuted modifications. Across the covered refactoring instances, 51.8\% of refactoring instance-workload comparisons exhibit statistically significant energy changes. Both energy increases and decreases occur under every workload, with neither direction predominating. Thus, although refactoring preserves functional behaviour, it often impact energy consumption, and its impact cannot be assumed to be energy-neutral or consistently beneficial.
\end{findingbox}

\begin{figure}[t]
  \centering
  \includegraphics[width=0.6\linewidth]
  {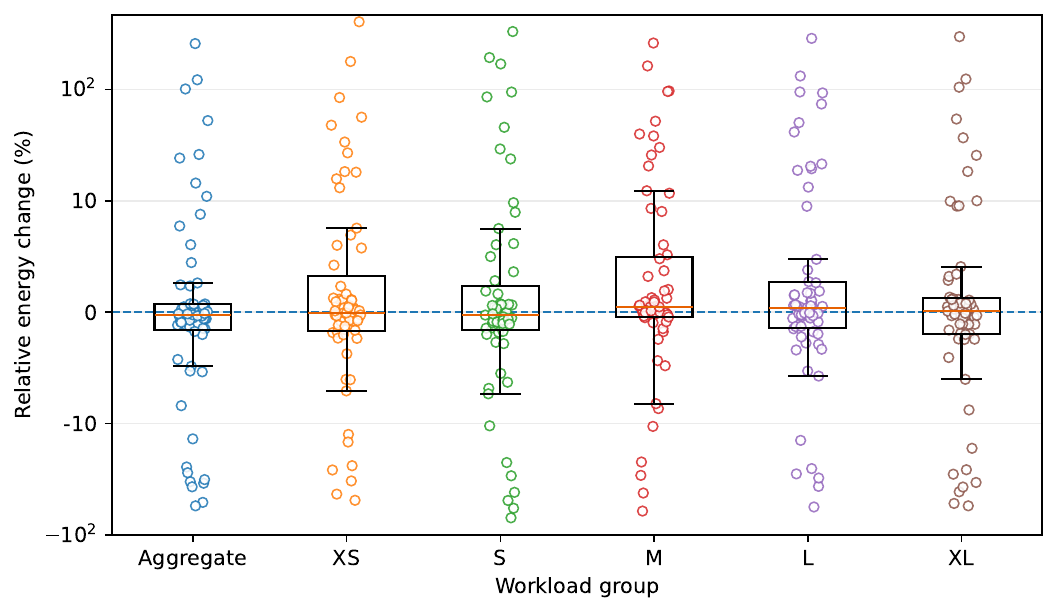}
  \caption{Item-level relative energy changes across workload groups.}
  \label{fig:rq1_micro_change_distribution}
\end{figure}

\paragraph{Magnitude and effect sizes of energy changes.}
Figure~\ref{fig:rq1_micro_change_distribution} presents boxplots overlaid with
item-level relative energy changes for each workload group. We use absolute
relative changes of 10\%, 50\%, and 100\% as descriptive magnitude thresholds.
Of the 384 item--workload comparisons, 107 (27.9\%) change by at least 10\%.
All 107 are statistically significant and span 26 of the
\microeligiblenumber{} eligible items. Their directional distributions differ
across workloads: Aggregate is balanced between increases and decreases (9/9),
and XL is nearly balanced (9/8), whereas M (13/5) and L (13/6) are weighted
toward increases. This directional imbalance strengthens at larger magnitudes:
increases account for 34 of the 42 comparisons that change by at least 50\%.
Moreover, 21 comparisons spanning five refactoring items and all six workload
groups increase mean energy consumption by at least 100\%. Thus, substantial
changes occur under every workload, but energy increases dominate the
higher-magnitude portion of the distributions.

The effect-size results provide complementary evidence
(Table~\ref{tab:rq1_micro_significance_effect}). All 199 statistically
significant comparisons have non-negligible Cliff's-$\delta$ values: 17
(8.5\%) are small, 20 (10.1\%) are medium, and 162 (81.4\%) are large. In
contrast, 129 of the 185 non-significant comparisons (69.7\%) have negligible
effects, and only three (1.6\%) have medium or large effects. Therefore, the
statistically significant energy differences are generally accompanied by
substantial separation between the before and after measurements.

\begin{table}[t]
  \centering
  \caption{Statistical significance and Cliff's-$\delta$ magnitude across the
  384 item--workload comparisons.}
  \label{tab:rq1_micro_significance_effect}
  \resizebox{0.6\linewidth}{!}{%
    \begin{tabular}{lrrrr}
      \toprule
      \textbf{Statistical result} &
      \textbf{Negligible} &
      \textbf{Small} &
      \textbf{Medium} &
      \textbf{Large} \\
      \midrule
      Significant ($n=199$) &
      0 (0.0\%) &
      17 (8.5\%) &
      20 (10.1\%) &
      162 (81.4\%) \\
      Not significant ($n=185$) &
      129 (69.7\%) &
      53 (28.6\%) &
      2 (1.1\%) &
      1 (0.5\%) \\
      \bottomrule
    \end{tabular}%
  }
\end{table}

\begin{findingbox}
Energy changes induced by refactoring are often substantial rather than marginal. More than one-quarter (27.9\%) of item--workload comparisons change energy consumption by at least 10\%, and these changes are consistently statistically significant with predominantly large effect sizes. Although both energy increases and decreases occur, larger-magnitude changes are increasingly dominated by energy increases, with all observed changes of at least 100\% corresponding to increased energy consumption. These results indicate that refactoring can produce practically meaningful energy effects and that substantial energy regressions, while not universal, represent a non-negligible risk.
\end{findingbox}

\paragraph{Within-item workload sensitivity.}
To directly examine whether workload choice changes the inferred energy impact
of the same refactoring item, we compare each eligible item's increase,
decrease, or unchanged classification across the five input-size workloads
(XS--XL). An item is considered workload-sensitive when it does not receive
the same classification under all five workloads. Under this definition, 29
of the \microeligiblenumber{} eligible items (45.3\%) are workload-sensitive.
Nineteen items vary only between a statistically significant change and an
unchanged classification, without switching between increase and decrease.
More importantly, ten items (15.6\%) are classified as significant increases
under at least one input size and significant decreases under another.

We also calculate, for each item, the range between its maximum and minimum
relative energy changes across XS--XL. Twenty-one items (32.8\%) have a range
of at least 10 percentage points. These within-item differences demonstrate
that measuring a refactoring under only one workload may fail to detect an
effect observed under another workload or may produce the opposite directional
conclusion.

\begin{findingbox}
Workload choice affects the inferred energy impact of individual refactorings.
Across XS--XL, 29 of the \microeligiblenumber{} eligible items (45.3\%) change
their increase, decrease, or unchanged classification. This includes ten items
(15.6\%) that switch between significant energy increases and decreases.
Moreover, 21 items (32.8\%) exhibit a relative-energy-change range of at least
10 percentage points. Thus, a single workload may miss or even reverse the
observed energy effect of a refactoring.
\end{findingbox}

Overall, the \microbenchmark{} results show that refactoring can produce
statistically reliable energy increases and decreases across all workload
groups. Although changes occur in both directions, energy increases become
increasingly prominent among the largest observed differences. The within-item
comparisons further show that workload selection can change whether the same
refactoring item is classified as an increase, decrease, or unchanged. These
controlled results demonstrate both the potential energy impact of individual
refactorings and the importance of measuring them under diverse workloads; we
next examine whether energy differences also arise in practical refactoring
commits.

\subsubsection{Energy Impact in \practicalbenchmark{}}
\label{subsubsec:result_rq1_practicalbenchmark}

\paragraph{Analysis population and commit composition.}
The RQ1 analysis covers 481 unique refactoring commits from 430 GitHub
projects. For each commit, we evaluate the energy impact using 30 matched
before/after energy measurements, yielding 14,430 matched measurement pairs (481 commits $\times$ 30 measurements).
Table~\ref{tab:rq1_practical_composition_summary} summarizes the analysis
population and the composition of the commits. The composition of these commits differs substantially from the isolated
transformations in \microbenchmark{}. Among the 479 commits with RefactoringMiner
output, 56 (11.7\%) contain exactly one detected refactoring operation, whereas 289
(60.3\%) contain multiple operations. Similarly, 89 (18.6\%) contain exactly
one distinct refactoring type, whereas 256 (53.4\%) contain multiple distinct types.
RefactoringMiner records no supported operation for the remaining 134 commits
(28.0\%). Among the 345 commits with at least one detected refactoring operation, 83.8\%
contain multiple operations and 74.2\% contain multiple distinct types.
Together with the 278 commits that modify multiple production Java files,
these results show that commit-level measurements frequently capture bundles
of transformations rather than a single isolated refactoring.

\begin{table}[t]
  \centering
  \caption{Population and commit-composition statistics for
  \practicalbenchmark{}.}
  \label{tab:rq1_practical_composition_summary}
  \begin{adjustbox}{width=0.6\linewidth}
    \begin{tabular}{@{}lr@{}}
      \toprule
      \textbf{Characteristic} & \textbf{Value} \\
      \midrule
      Energy-evaluable commits & 481 \\
      GitHub projects & 430 \\
      Measurement pairs per commit & 30 \\
      Commits with RefactoringMiner output & 479 \\
      Commits with at least one detected refactoring & 345 \\
      Detected refactoring per analyzed commit & Mean 10; median 3 \\
      Distinct detected refactoring per analyzed commit & Mean 3; median 2 \\
      Commits modifying multiple production Java files & 278 (57.8\%) \\
      \bottomrule
    \end{tabular}
  \end{adjustbox}
\end{table}

\paragraph{Frequency and direction of energy changes.}
As in the \microbenchmark{} analysis, a commit is statistically significant
when its paired Wilcoxon signed-rank test has $p<0.05$. A significant commit is
classified as an increase or decrease using the sign of its mean paired
difference. Unchanged denotes the absence of a statistically significant
difference under the executed project test-suite workload rather than equality
between the before and after measurements.

Table~\ref{tab:rq1_practical_outcome_summary} summarizes the commit-level
classifications. Thirty-six of the 481 commits (7.5\%) exhibit a statistically significant energy difference. Fifteen
commits (3.1\%) increase energy consumption, 21 (4.4\%) decrease it, and 445
(92.5\%) are unchanged. The 36 significant commits come from 36 different
projects, so the result is not driven by multiple significant commits from one
project. Among these commits, increases account for 41.7\% and decreases for
58.3\%; a two-sided exact binomial test finds no directional imbalance
($p=0.405$).

\begin{table}[t]
  \centering
  \caption{Commit-level energy-change classifications in
  \practicalbenchmark{}.}
  \label{tab:rq1_practical_outcome_summary}
  \begin{adjustbox}{width=0.28\linewidth}
    \begin{tabular}{@{}lrr@{}}
      \toprule
      \textbf{Classification} & \textbf{$n$} & \textbf{\%} \\
      \midrule
      Significant increase & 15 & 3.1 \\
      Unchanged & 445 & 92.5 \\
      Significant decrease & 21 & 4.4 \\
      \midrule
      Total & 481 & 100.0 \\
      \bottomrule
    \end{tabular}
  \end{adjustbox}
\end{table}

\begin{findingbox}
Real-world refactoring commits typically comprise multiple co-occurring refactoring operations and often modify multiple source files, reflecting the complexity of practical software evolution. Under these realistic conditions, statistically significant energy changes are relatively uncommon, occurring in only 7.5\% of commits. When they do occur, both energy increases and decreases are observed, with neither direction predominating. These results suggest that most real-world refactoring commits are effectively energy-neutral under the evaluated workloads, but a non-negligible subset can still produce measurable energy changes.
\end{findingbox}

\begin{figure}[t]
  \centering
  \includegraphics[width=0.6\linewidth]
  {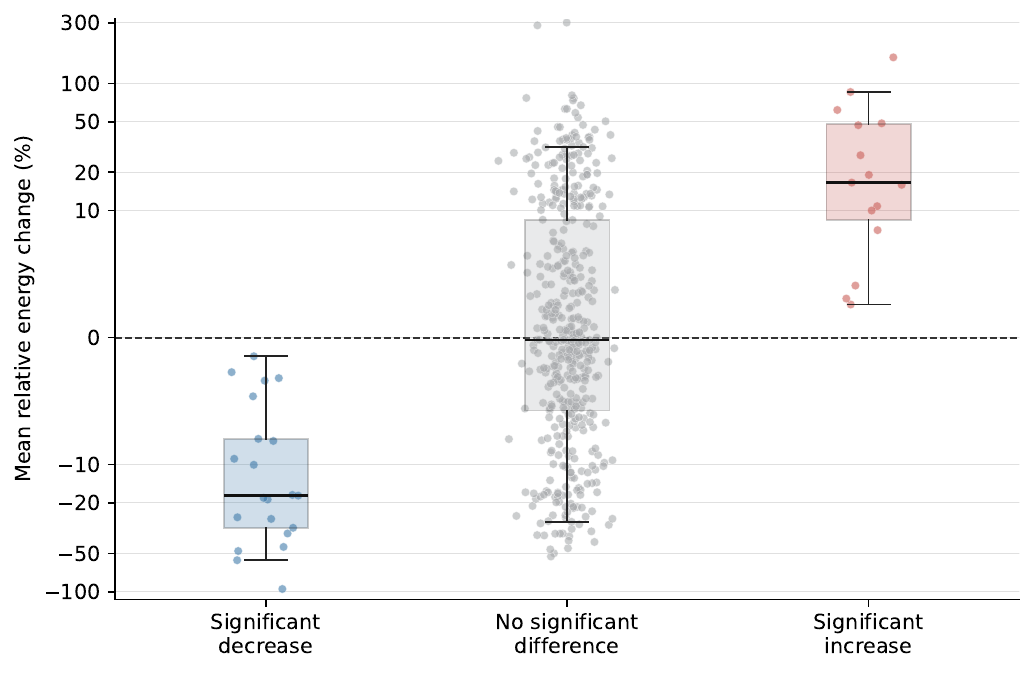}
  \caption{Distributions of commit-level mean relative energy changes by
  statistical classification. Dots represent individual commits, and boxes
  summarize medians and quartiles. The vertical axis uses a symmetric
  logarithmic scale with a linear region from $-10\%$ to $10\%$.}
  \Description{Commit-level mean relative energy changes for 21 significant
  decreases, 445 unchanged commits, and 15 significant increases.}
  \label{fig:rq1_practical_change_distribution}
\end{figure}

\paragraph{Magnitude and effect sizes of energy changes.}
Figure~\ref{fig:rq1_practical_change_distribution} shows the distribution of
mean relative energy changes. Across all 481 commits, the median change is
$-0.25\%$ (first quartile: $-6.35\%$; third quartile: $9.73\%$), with values
ranging from $-95.21\%$ to $301.89\%$. The wide range among unchanged commits
also shows why the difference between sample means should be considered
together with repeated-measurement statistics. Among the 36 statistically significant commits, the median increase is
$16.61\%$, while the median decrease is $-17.52\%$. Using the same descriptive
magnitude thresholds as in the \microbenchmark{} analysis, 24 of the 36 commits
(66.7\%) change by at least 10\% (11 increases and 13 decreases), five (13.9\%)
change by at least 50\% (three increases and two decreases), and one (2.8\%)
increases by at least 100\%.

Table~\ref{tab:rq1_practical_significance_effect} summarizes the conventional
Cliff's-$\delta$ effect sizes. Thirty of the 36 significant commits (83.3\%)
have non-negligible effects: 16 are small, ten are medium, and four are large.
The remaining six have negligible effects. Medium or large effects are strongly
concentrated among statistically significant commits: 14 of 36 significant
commits (38.9\%), compared with two of 445 unchanged commits (0.4\%). These results show that many of the detected
energy differences are supported by both substantial relative changes.

\begin{table}[t]
  \centering
  \caption{Statistical significance and Cliff's-$\delta$ magnitude across the
  481 refactoring commits.}
  \label{tab:rq1_practical_significance_effect}
  \begin{adjustbox}{width=0.6\linewidth}
    \begin{tabular}{@{}lrrrr@{}}
      \toprule
      \textbf{Statistical result} &
      \textbf{Negligible} &
      \textbf{Small} &
      \textbf{Medium} &
      \textbf{Large} \\
      \midrule
      Significant ($n=36$) &
      6 (16.7\%) &
      16 (44.4\%) &
      10 (27.8\%) &
      4 (11.1\%) \\
      Unchanged ($n=445$) &
      361 (81.1\%) &
      82 (18.4\%) &
      2 (0.4\%) &
      0 (0.0\%) \\
      \bottomrule
    \end{tabular}
  \end{adjustbox}
\end{table}

\begin{findingbox}
The statistically significant commit-level energy differences vary in
magnitude, with most commits exhibiting non-negligible effect sizes.
Among the 36 significant commits, 24 (66.7\%) change mean energy
consumption by at least 10\%, including five (13.9\%) with changes of
at least 50\% and one (2.8\%) with an increase of more than 100\%.
Regarding effect size, 30 commits (83.3\%) have non-negligible
Cliff's-$\delta$ values, including 14 (38.9\%) with medium or large
effects.
\end{findingbox}

Overall, the \practicalbenchmark{} results extend the controlled evidence from
\microbenchmark{} to complete refactoring commits drawn from real projects.
Although most commits are classified as unchanged under their test-suite
workloads, statistically significant increases and decreases occur and can be
substantial. Because these commits frequently combine multiple operations,
distinct refactoring types, and production files, their measured energy impact
reflects the complete change rather than the isolated effect of any single
operation. We next examine which refactoring types and co-occurrence patterns
are associated with these energy differences.

\subsection{RQ2: Which refactoring types are more likely to influence energy consumption?}
\label{subsec:result_rq2}

RQ1 shows that refactoring can produce statistically significant increases and
decreases in energy consumption. RQ2 examines whether these differences are
concentrated among particular refactoring types and, when they occur, how large
they are. We compare refactoring types along two complementary dimensions. The
first is the observed frequency of statistically significant before/after
differences under the paired Wilcoxon signed-rank criterion ($p<0.05$) used in
RQ1. The second is the magnitude of these differences: for evaluations that
exhibit a statistically significant difference, we further analyze the direction
and mean relative change in energy consumption, together with the
Cliff's-$\delta$ effect size. The frequency dimension shows how recurrently an
energy difference is observed for a type, whereas the relative change and effect
size characterize the extent of the detected difference. We therefore interpret
these dimensions together when determining which refactoring types are more
likely to influence energy consumption.

In \microbenchmark{}, each benchmark item isolates a target refactoring type,
allowing its energy behavior to be examined in a controlled setting. For each
type, we analyze the frequency and direction of statistically significant
differences across the six workload groups---Aggregate, XS, S, M, L, and
XL---and characterize their relative changes and Cliff's-$\delta$ effect sizes.
Aggregate represents the combined workload, whereas XS--XL represent individual
input-size workloads. We retain this distinction when interpreting the results,
but examine all six groups together to provide a unified view of each
refactoring type's energy behavior across the evaluated workloads.

In \practicalbenchmark{}, the measured energy difference belongs to the complete
commit. We therefore distinguish commits containing exactly one distinct
detected refactoring type from those containing multiple distinct types. For
single-type commits, we examine the frequency, direction, and relative-change
magnitude of the commit-level differences according to the sole detected type.
Although such a commit may contain multiple operations of that type, it provides
more direct type-specific evidence because no other detected refactoring type
co-occurs in the commit. For multi-type commits, we examine whether individual
types and sufficiently supported co-occurrence patterns are more prevalent among
commits with statistically significant increases or decreases than among
unchanged commits. We also characterize the relative-change magnitudes of the
corresponding significant commits. These multi-type results are interpreted as
associations because the contribution of one type cannot be isolated from those
of the other co-occurring types. Accordingly, \microbenchmark{} provides
controlled evidence for isolated types, single-type practical commits provide
type-focused evidence in real-world contexts, and multi-type practical commits
provide evidence about individual-type and co-occurrence associations.

\subsubsection{Refactoring-Type Effects in \microbenchmark{}}
\label{subsubsec:result_rq2_microbenchmark}

\paragraph{Analysis population and type-level measures.}
We reuse the 64 \microbenchmark{} items included in the RQ1 analysis, each of
which has valid before/after energy measurements for Aggregate, XS, S, M, L,
and XL. Four items target Extract Function/Method, while each of the remaining
60 items represents a different refactoring type; the analysis therefore
covers 61 refactoring types. For each item, we conduct one before/after
statistical comparison in each of the six workload groups, yielding
$64 \times 6 = 384$ item--workload comparisons. Each comparison is based on 30
paired before/after measurements; these measurements support the statistical
test but are not counted as separate item--workload comparisons. For each
refactoring type, we calculate the observed frequency of statistically
significant energy differences as the proportion of its item--workload
comparisons that yield a significant result, reporting increases and decreases
separately. The denominator is therefore six comparisons for each single-item
type and 24 comparisons for Extract Function/Method. We use mean relative
energy change and Cliff's-$\delta$ to characterize the magnitude of the
observed differences. Aggregate is a combined workload and XS--XL are
individual input-size workloads, but all six are examined together as the
tested workload groups.

\begin{figure*}[t]
  \centering
  \includegraphics[width=0.8\textwidth]
  {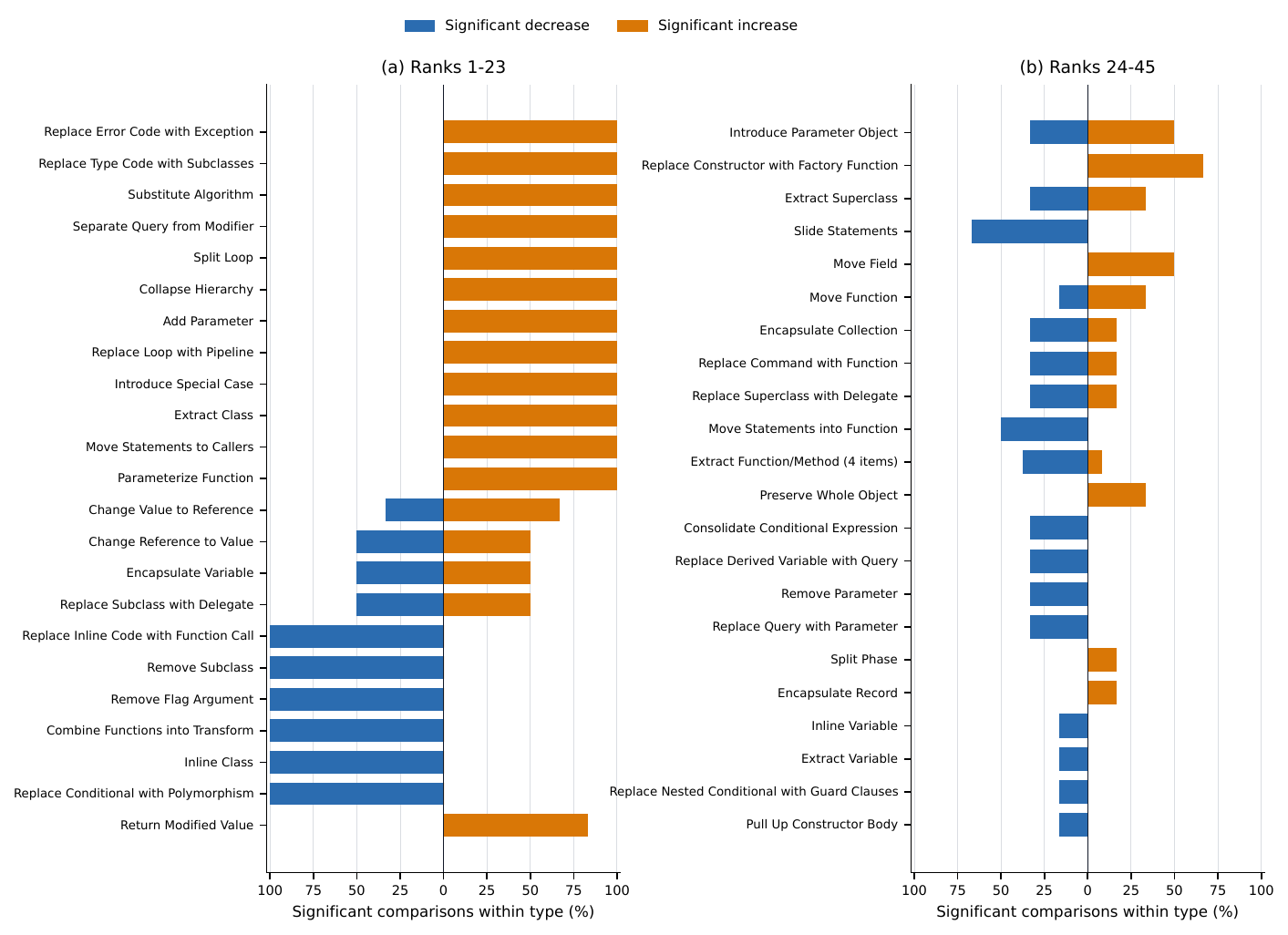}
  \caption{Observed frequency and direction of statistically significant
  energy differences by refactoring type in \microbenchmark{}. Only the 45
  types with at least one significant item--workload comparison are shown. For
  each type, the bars show the percentages of its comparisons classified as
  significant increases or decreases across Aggregate, XS, S, M, L, and XL.
  Types are ordered by the sum of these percentages and split into two panels
  for readability.}
  \Description{Diverging bars for the 45 refactoring types with at least one
  significant item--workload comparison. Twenty-two types exhibit significant
  differences under all six workload groups. Consistently increasing,
  consistently decreasing, and mixed-direction types are present.}
  \label{fig:rq2_micro_type_frequency}
\end{figure*}

\paragraph{Frequency and direction across workload groups.}
Figure~\ref{fig:rq2_micro_type_frequency} ranks the 45 refactoring types that
exhibit at least one statistically significant item--workload comparison by
the proportion of their comparisons classified as significant increases or
decreases. These 45 types constitute 73.8\% of the 61 types; the other 16
types (26.2\%) exhibit no significant difference and are omitted from the
figure. Among the 45 types shown, 18 (40.0\%) exhibit only increases, 16
(35.6\%) exhibit only decreases, and 11 (24.4\%) exhibit both directions.

However, only 22 of the 61 types (36.1\%) exhibit a statistically significant
energy difference across all six workload groups. Among these 22 types, 12
(54.5\%) consistently increase energy
consumption, including Replace Error Code with Exception, Replace Type Code
with Subclasses, Substitute Algorithm, and Replace Loop with Pipeline. Six
(27.3\%) consistently decrease energy consumption, including Replace Inline
Code with Function Call, Remove Subclass, and Remove Flag Argument. The
remaining four (18.2\%)---Change Value to Reference, Change Reference to Value,
Encapsulate Variable, and Replace Subclass with Delegate---exhibit significant
differences in all six workload groups but include both increases and
decreases.

\begin{findingbox}
Statistically significant energy differences occur for most refactoring types,
but only a smaller subset exhibits such differences across all six workload
groups. Of the 61 types, 45 (73.8\%) exhibit at least one significant
item--workload comparison, whereas 22 (36.1\%) exhibit a significant difference
across all six workload groups.
Among these 22 types, 12 (54.5\%) consistently increase energy consumption, six
(27.3\%) consistently decrease it, and four (18.2\%) exhibit both directions.
The other 16 types (26.2\%) exhibit no significant difference.
\end{findingbox}

\paragraph{Magnitude and workload dependence of refactoring-type effects.}

\begin{figure*}[t]
  \centering
  \includegraphics[width=0.8\textwidth]
  {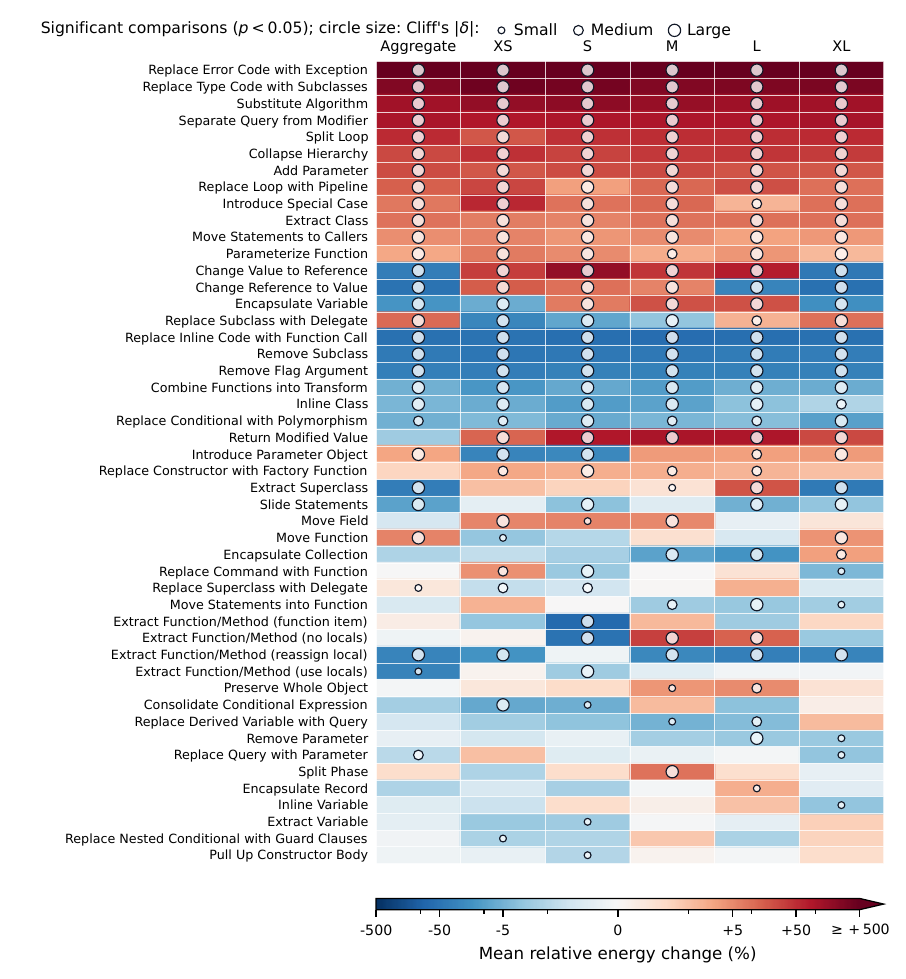}
  \caption{Mean relative energy change across Aggregate, XS, S, M, L, and XL
  for all 45 refactoring types with at least one statistically significant
  comparison in \microbenchmark{}. The four Extract Function/Method items are
  displayed as separate rows, resulting in 48 benchmark-item rows. Red denotes
  higher energy after refactoring and blue denotes lower energy. Circles mark
  statistically significant comparisons, with their sizes representing small,
  medium, and large Cliff's-$\delta$ effects.}
  \Description{Heatmap showing all 45 refactoring types with at least one
  statistically significant comparison across six workload groups. Both
  directionally consistent and workload-dependent patterns are visible.}
  \label{fig:rq2_micro_type_heatmap}
\end{figure*}

Figure~\ref{fig:rq2_micro_type_heatmap} complements the frequency view by
showing the mean relative energy change and Cliff's-$\delta$ magnitude for all
45 refactoring types with at least one statistically significant comparison.
The four Extract Function/Method items are shown separately so that their
within-type variation is not hidden by aggregation. Of the 384 item--workload comparisons defined above, 199 (51.8\%) yield a
statistically significant difference: 113 increases and 86 decreases. Among
these 199 comparisons, 162 (81.4\%) have a large Cliff's-$\delta$ effect, 20
(10.1\%) have a medium effect, and 17 (8.5\%) have a small effect; none has a
negligible effect. The 22 types that differ across all six workload groups
account for 132 of these comparisons. Within this subset, 124 (93.9\%) have a
large effect and the remaining eight (6.1\%) have a medium effect.
Table~\ref{tab:rq2_micro_all_workloads} reports the direction, median relative
change, range, and number of large effects for these 22 types.

\begin{table*}[t]
  \centering
  \caption{The 22 refactoring types with statistically significant differences
  across all six workload groups. Inc./dec. reports the numbers of increases
  and decreases among the six comparisons. Median and range summarize the six
  mean relative energy changes. Large $|\delta|$ reports how many comparisons
  have a large Cliff's-$\delta$ effect.}
  \label{tab:rq2_micro_all_workloads}
  \small
  \setlength{\tabcolsep}{4pt}
  \begin{adjustbox}{max width=0.7\textwidth}
    \begin{tabular}{@{}lcrrc@{}}
      \toprule
      \textbf{Refactoring type} &
      \textbf{Inc./dec.} &
      \textbf{Median $\Delta E$} &
      \textbf{Range of $\Delta E$} &
      \textbf{Large $|\delta|$} \\
      \midrule
      \multicolumn{5}{@{}l}{\textit{Consistent increases (12 types)}} \\
      \addlinespace[1pt]
      Replace Error Code with Exception
        & 6/0 & $+13{,}022.16\%$
        & $[+12{,}216.65\%,+13{,}778.64\%]$ & 6/6 \\
      Replace Type Code with Subclasses
        & 6/0 & $+293.60\%$ & $[+258.84\%,+406.74\%]$ & 6/6 \\
      Substitute Algorithm
        & 6/0 & $+147.80\%$ & $[+122.37\%,+193.92\%]$ & 6/6 \\
      Separate Query from Modifier
        & 6/0 & $+95.57\%$ & $[+84.76\%,+105.04\%]$ & 6/6 \\
      Split Loop
        & 6/0 & $+51.21\%$ & $[+18.10\%,+54.35\%]$ & 6/6 \\
      Collapse Hierarchy
        & 6/0 & $+37.59\%$ & $[+26.11\%,+47.88\%]$ & 6/6 \\
      Add Parameter
        & 6/0 & $+21.62\%$ & $[+18.31\%,+25.82\%]$ & 6/6 \\
      Replace Loop with Pipeline
        & 6/0 & $+13.38\%$ & $[+2.83\%,+27.00\%]$ & 6/6 \\
      Introduce Special Case
        & 6/0 & $+9.25\%$ & $[+1.52\%,+56.49\%]$ & 5/6 \\
      Extract Class
        & 6/0 & $+8.92\%$ & $[+6.06\%,+9.97\%]$ & 6/6 \\
      Move Statements to Callers
        & 6/0 & $+4.04\%$ & $[+2.64\%,+6.01\%]$ & 6/6 \\
      Parameterize Function
        & 6/0 & $+3.07\%$ & $[+1.35\%,+7.55\%]$ & 5/6 \\
      \addlinespace[3pt]
      \multicolumn{5}{@{}l}{\textit{Consistent decreases (6 types)}} \\
      \addlinespace[1pt]
      Replace Inline Code with Function Call
        & 0/6 & $-55.64\%$ & $[-61.26\%,-49.03\%]$ & 6/6 \\
      Remove Subclass
        & 0/6 & $-39.33\%$ & $[-43.14\%,-33.90\%]$ & 6/6 \\
      Remove Flag Argument
        & 0/6 & $-29.40\%$ & $[-32.85\%,-28.38\%]$ & 6/6 \\
      Combine Functions into Transform
        & 0/6 & $-6.44\%$ & $[-12.54\%,-5.36\%]$ & 6/6 \\
      Inline Class
        & 0/6 & $-5.16\%$ & $[-10.50\%,-1.09\%]$ & 5/6 \\
      Replace Conditional with Polymorphism
        & 0/6 & $-4.82\%$ & $[-8.78\%,-3.33\%]$ & 2/6 \\
      \addlinespace[3pt]
      \multicolumn{5}{@{}l}{\textit{Mixed directions (4 types)}} \\
      \addlinespace[1pt]
      Change Value to Reference
        & 4/2 & $+36.80\%$ & $[-41.12\%,+169.92\%]$ & 6/6 \\
      Change Reference to Value
        & 3/3 & $-9.70\%$ & $[-52.31\%,+15.78\%]$ & 6/6 \\
      Encapsulate Variable
        & 3/3 & $+0.73\%$ & $[-16.72\%,+20.58\%]$ & 6/6 \\
      Replace Subclass with Delegate
        & 3/3 & $-0.38\%$ & $[-23.97\%,+10.98\%]$ & 5/6 \\
      \bottomrule
    \end{tabular}
  \end{adjustbox}
\end{table*}

The magnitude of all-workload differences varies substantially even within
the directionally consistent groups. Across the 12 consistently increasing
types, the median relative change ranges from $+3.07\%$ for Parameterize
Function to $+13{,}022.16\%$ for Replace Error Code with Exception. The next-largest medians are $+293.60\%$ for Replace Type
Code with Subclasses. Across the six
consistently decreasing types, the medians range from $-4.82\%$ for Replace
Conditional with Polymorphism to $-55.64\%$ for Replace Inline Code with
Function Call. The four mixed-direction types further show that significance
across all six workload groups does not imply a stable direction. For example,
Change Value to Reference has four increases and two decreases, ranging from
$-41.12\%$ to $+169.92\%$. 

The other 23 types in the heatmap exhibit significant differences only for a
subset of their comparisons, revealing additional workload dependence. Return
Modified Value is unchanged under Aggregate but increases energy under all five
input-size workloads, with changes from $+13.12\%$ to $+97.12\%$. Introduce
Parameter Object decreases energy under XS and S, increases it under Aggregate,
L, and XL, and is unchanged under M. The four Extract Function/Method items
provide within-type evidence: 11 of their 24 comparisons are significant,
comprising two increases and nine decreases, with significant changes ranging
from $-70.46\%$ to $+30.20\%$. Therefore, both the executed workload and the
concrete benchmark item can alter whether an energy difference is detected, its
direction, and its magnitude.

\begin{findingbox}
Refactoring-type effects are often large but remain workload-dependent: 162 of
the 199 statistically significant item--workload comparisons (81.4\%) have
large Cliff's-$\delta$ effects, yet effect magnitudes vary across workload
groups. Refactoring type therefore indicates the potential for energy impact but does not determine a fixed energy outcome.
\end{findingbox}

\paragraph{Overall refactoring-type patterns.}
Taken together, the controlled results reveal two complementary type-level
patterns. First, 22 of the 61 types (36.1\%) exhibit statistically significant
energy differences across all six workload groups, and 18 of these 22 (81.8\%)
maintain the same direction. Three consistently increasing types---Replace Error
Code with Exception, Replace Type Code with Subclasses, and Substitute
Algorithm---have the three largest median increases, whereas three consistently
decreasing types---Replace Inline Code with Function Call, Remove Subclass, and
Remove Flag Argument---have the three largest median decreases. All six
comparisons for each of these six types have large Cliff's-$\delta$ effects. However, these six refactoring types are not supported by RefactoringMiner, preventing their automatic identification in practical commits. This limitation does not imply that the refactorings are absent from real-world development; rather, they are often realized through complex sequences of coordinated edits that are difficult for existing refactoring detectors to recognize as instances of a single high-level refactoring. As a result, we cannot isolate their occurrences in \practicalbenchmark{} to verify whether they exhibit the same directional energy behavior observed in \microbenchmark{}. Accordingly, we interpret these findings as evidence for the controlled transformations evaluated in our study, rather than as universal properties of these refactoring types.
Second, the remaining evidence is more workload-specific. Among
the 45 types with at least one significant comparison, 23 differ only under
selected workload groups, and 11 exhibit both increases and decreases across
their significant comparisons. Refactoring type can therefore help prioritize
which transformations to evaluate for energy impact, but it cannot by itself
predict a fixed direction or magnitude across workloads. We next examine whether
similar type-level patterns appear among the refactoring types detected in
practical commits, separating single-type commits from those containing multiple
refactoring types.

\subsubsection{Refactoring Patterns Associated with Energy Differences in
\practicalbenchmark{}}
\label{subsubsec:result_rq2_practicalbenchmark}

\paragraph{Analysis population and commit-level refactoring composition.}
We begin with the 481 \practicalbenchmark{} commits analyzed in RQ1, drawn
from 430 projects and each containing 30 paired before/after energy
measurements. RefactoringMiner successfully analyzes 479 of these commits
(99.6\%). Since RefactoringMiner detects only a predefined set of supported
refactoring types,\footnote{\href{https://github.com/tsantalis/RefactoringMiner/blob/master/documentation/history.md\#refactoring-support-history}
{RefactoringMiner refactoring support history.}} it records no operation in
134 of the 479 commits (28.0\%). These commits may contain refactorings outside
the supported set or instances that the tool does not detect. We report these
commits for transparency but exclude them from the type- and pattern-level
analyses because they provide no detected refactoring type. For
the remaining two commits, RefactoringMiner exceeds its default 90-second time limit and returns no analysis result. These two timeout commits are therefore excluded. The resulting RQ2 analysis population therefore contains 345 commits from 318 projects, each with at least one detected refactoring operation.

We distinguish the number of refactoring operations from the number of
distinct refactoring types in a commit. Each detected occurrence counts as one
operation, whereas a type is counted once regardless of how many times it
occurs. Accordingly, a \emph{single-type commit} contains exactly one distinct
detected type but may contain multiple operations of that type, while a
\emph{multi-type commit} contains at least two distinct detected types. Of the
345 commits, 89 (25.8\%) are single-type commits and 256 (74.2\%) are
multi-type commits. In particular, 33 of the 89 single-type commits (37.1\%)
apply their sole type more than once, confirming that ``single-type'' does not
mean ``single-operation''. Because the subsequent individual-type analysis relies on the single-type commits, we also assess how broadly they represent the detected refactoring types. The 89 single-type commits span 26 types, whereas
the multi-type commits and the complete RQ2 population span all 90 detected
types.

Table~\ref{tab:rq2_practical_composition} relates this composition to the
commit-level energy classifications established in RQ1. Among the 345
included commits, 12 (3.5\%) exhibit a statistically significant energy
increase, 14 (4.1\%) exhibit a statistically significant decrease, and 319
(92.5\%) are unchanged (i.e., no statistically significant difference is
detected). A significant difference is observed in 10 of the 89 single-type
commits (11.2\%) and 16 of the 256 multi-type commits (6.3\%). This difference
is not statistically significant (Fisher's exact test:
$\mathrm{OR}=1.89$, exact 95\% CI: $[0.74,4.65]$, $p=0.160$), providing no
statistically significant evidence that the number of distinct detected types
is associated with a greater likelihood of observing an energy difference.
Nevertheless, the predominance of multi-type commits means that most practical
outcomes cannot be attributed to one isolated refactoring type. We therefore
next examine individual refactoring types and recurring combinations of types.

\begin{table*}[t]
  \centering
  \caption{Commit-level refactoring composition and energy classifications in
  \practicalbenchmark{}. ``Total distinct types'' reports the number of distinct
  refactoring types detected across all commits in each group. Operations and
  distinct types are reported per commit as median [$Q_1$--$Q_3$], where $Q_1$
  and $Q_3$ are the 25th and 75th percentiles. Outcome percentages are calculated
  within each commit group. ``Unchanged'' denotes no statistically significant
  energy difference.}
  \label{tab:rq2_practical_composition}
  \small
  \setlength{\tabcolsep}{4pt}
  \begin{adjustbox}{max width=0.98\textwidth}
    \begin{tabular}{@{}lrrrrrrrr@{}}
      \toprule
      \textbf{Commit group} &
      \textbf{Commits} &
      \textbf{Projects} &
      \shortstack{\textbf{Operations}\\\textbf{median [$Q_1$--$Q_3$]}} &
      \shortstack{\textbf{Distinct types}\\\textbf{median [$Q_1$--$Q_3$]}} &
      \shortstack{\textbf{Total distinct}\\\textbf{types}} &
      \textbf{Increase} &
      \textbf{Decrease} &
      \textbf{Unchanged} \\
      \midrule
      No detected type
        & 134 & 121 & 0 [0--0] & 0 [0--0] & 0
        & 3 (2.2\%) & 7 (5.2\%) & 124 (92.5\%) \\
      Single-type
        & 89 & 86 & 1 [1--2] & 1 [1--1] & 26
        & 6 (6.7\%) & 4 (4.5\%) & 79 (88.8\%) \\
      Multi-type
        & 256 & 238 & 9 [4--20] & 4 [3--8] & 90
        & 6 (2.3\%) & 10 (3.9\%) & 240 (93.8\%) \\
      \midrule
      \textbf{RQ2 analysis population}
        & \textbf{345} & \textbf{318} & \textbf{5 [2--14]}
        & \textbf{3 [1--6]} & \textbf{90}
        & \textbf{12 (3.5\%)} & \textbf{14 (4.1\%)}
        & \textbf{319 (92.5\%)} \\
      \bottomrule
    \end{tabular}
  \end{adjustbox}
\end{table*}

\paragraph{Individual types and co-occurring patterns associated with energy
differences.}
We first analyze single-type commits to determine whether any individual
refactoring type exhibits a repeatable energy-increase or energy-decrease tendency
in real-world commits. Each such commit contains only one distinct refactoring type
detected by RefactoringMiner, although it may contain multiple instances of that
type. This reduces the ambiguity introduced by co-occurring types and provides the
clearest practical evidence for individual-type effects. Before examining their
energy outcomes, we assess the diversity of the single-type group and the number of
commits available for each type. These statistics indicate whether the practical
benchmark provides sufficiently broad and repeated evidence to support general
conclusions about particular refactoring types.

The 89 single-type commits cover 26 of the 90 refactoring types detected in the
full \practicalbenchmark{} population. However, the evidence is unevenly
distributed: 13 of the 26 types occur in only one single-type commit, whereas only
Inline Variable (20 commits) and Extract Method (14 commits) occur in at least ten
commits. Thus, the single-type group covers a range of refactoring types, but the
limited number of observations for most types restricts our ability to establish
repeatable type-level energy tendencies.

We then examine whether any of these 26 types is repeatedly associated with
statistically significant energy increases or decreases. Only 10 of the 89
single-type commits exhibit a significant energy difference: six increases and
four decreases distributed across nine types. Inline Variable is the only type
represented by more than one significant commit, and both differences have
negligible Cliff's-$\delta$ magnitudes. Moreover, no individual type remains
associated with an energy increase or decrease after correcting the type-level
tests for multiple comparisons using the Benjamini--Hochberg procedure. Therefore,
the current single-type evidence does not support assigning a fixed energy
direction to any practical refactoring type.

We therefore focus the co-occurrence analysis on pairs and triples of types in
the 256 multi-type commits, without repeating a size-one presence analysis.
As Table~\ref{tab:rq1_practical_composition_summary} shows, the number of
distinct detected refactoring types per analyzed practical commit has a mean
of three and a median of two. Pairs and triples consequently represent the
most typical co-occurrence sizes, whereas larger combinations would be
increasingly sparse and project-specific. We treat each pattern as an
unordered set of distinct types and count it at most once per commit,
regardless of how many operations of its constituent types occur. This binary
definition matches the commit-level energy outcome and prevents commits with
many repeated operations from receiving disproportionate weight. We retain patterns occurring in at least five of the 256 multi-type commits. For each pair or triple, we compare commits containing
the pattern with the other multi-type commits using a two-sided Fisher's exact
test. We apply BH correction separately for pairs and triples and for increases
and decreases. This procedure evaluates 424 pairs and 1,142 triples for each
direction.

Table~\ref{tab:rq2_practical_type_pattern_associations} reports the three
strongest patterns in each pattern-size--outcome group. No
pair or triple associated with an energy increase survives correction; the
smallest adjusted value is $q=0.391$. In contrast, three pairs and one triple
remain associated with energy decreases. The triple \{Change Parameter Type,
Inline Variable, Remove Method Annotation\} occurs in six commits from six
projects, four of which exhibit a significant decrease (66.7\%), compared with
10 of the 256 multi-type commits overall (3.9\%). Its decrease rate is therefore
17.07 times the multi-type baseline.

The three surviving pairs are \{Inline Variable, Remove Method Annotation\}
($4/7$ decreases; $q=0.0168$), \{Change Parameter Type, Inline Variable\},
and \{Change Parameter Type, Remove Method Annotation\} (both $4/11$;
$q=0.0489$). The same four decrease commits support all four corrected
patterns. These rows therefore represent an overlapping co-occurrence signal,
not four independent effects. The next-ranked decrease triples also combine
parameter- or variable-type changes with annotation removal or code movement,
but they do not survive correction ($q=0.159$).

\begin{table*}[t]
  \centering
  \caption{Strongest recurring co-occurrence patterns for significant energy
  increases and decreases. For each pattern-size--outcome group, the three
  patterns with the smallest Fisher exact-test $p$-values are shown. All
  patterns occur in at least five multi-type commits. \emph{sup} is pattern
  support, \emph{out} is the number of commits with the target outcome, and
  \emph{conf} is \(\mathit{out}/\mathit{sup}\). Lift is calculated relative to the overall outcome rate among the 256 multi-type commits: 6/256 (2.3\%) for significant increases and 10/256 (3.9\%) for significant decreases. $q$ is the BH-adjusted value; bold values survive correction at 0.05.}
  \label{tab:rq2_practical_type_pattern_associations}
  \small
  \setlength{\tabcolsep}{4pt}
  \begin{adjustbox}{max width=0.8\textwidth}
    \begin{tabular}{@{}lrrrrr@{}}
      \toprule
      \textbf{Pattern} &
      \textbf{sup} &
      \textbf{out} &
      \textbf{conf} &
      \textbf{lift} &
      \textbf{$q$} \\
      \midrule
      \multicolumn{6}{@{}l}{\emph{Energy increase: pairs}} \\
      \{Change Attribute Type, Remove Thrown Exception Type\}
        & 5 & 2 & 40.0\% & 17.07 & 0.394 \\
      \{Inline Method, Rename Parameter\}
        & 5 & 2 & 40.0\% & 17.07 & 0.394 \\
      \{Add Parameter, Remove Thrown Exception Type\}
        & 6 & 2 & 33.3\% & 14.22 & 0.394 \\
      \addlinespace[2pt]
      \multicolumn{6}{@{}l}{\emph{Energy increase: triples}} \\
      \{Change Attribute Type, Change Return Type, Rename Method\}
        & 15 & 3 & 20.0\% & 8.53 & 0.391 \\
      \{Add Parameter, Change Return Type, Remove Thrown Exception Type\}
        & 5 & 2 & 40.0\% & 17.07 & 0.391 \\
      \{Add Parameter, Inline Method, Rename Method\}
        & 5 & 2 & 40.0\% & 17.07 & 0.391 \\
      \midrule
      \multicolumn{6}{@{}l}{\emph{Energy decrease: pairs}} \\
      \{Inline Variable, Remove Method Annotation\}
        & 7 & 4 & 57.1\% & 14.63 & \textbf{0.0168} \\
      \{Change Parameter Type, Inline Variable\}
        & 11 & 4 & 36.4\% & 9.31 & \textbf{0.0489} \\
      \{Change Parameter Type, Remove Method Annotation\}
        & 11 & 4 & 36.4\% & 9.31 & \textbf{0.0489} \\
      \addlinespace[2pt]
      \multicolumn{6}{@{}l}{\emph{Energy decrease: triples}} \\
      \{Change Parameter Type, Inline Variable,
        Remove Method Annotation\}
        & 6 & 4 & 66.7\% & 17.07 & \textbf{0.0198} \\
      \{Change Parameter Type, Change Variable Type, Move Attribute\}
        & 5 & 3 & 60.0\% & 15.36 & 0.159 \\
      \{Change Parameter Type, Change Variable Type,
        Remove Method Annotation\}
        & 5 & 3 & 60.0\% & 15.36 & 0.159 \\
      \bottomrule
    \end{tabular}
  \end{adjustbox}
\end{table*}

\begin{findingbox}
Real-world commits containing certain specific recurring combinations of refactoring types are substantially more likely to exhibit significant energy changes than multi-type commits overall. In our dataset, the strongest
evidence concerns energy reductions: 
commits containing the combination of Change Parameter Type, Inline Variable, and Remove Method Annotation show a 66.7\% decrease rate, representing a 17.1$\times$ increase over the 3.9\% baseline for all multi-type commits, whereas no recurring combination is significantly associated with energy increases after multiple-comparison correction. These results suggest that energy impact analysis should consider recurring refactoring combinations alongside individual refactoring types, motivating the identification of energy-aware refactoring patterns rather than just reasoning about isolated refactoring types.
\end{findingbox}

\paragraph{Magnitude and robustness of practical refactoring-pattern
associations.}
We next examine whether the four corrected associations correspond to broader
energy reductions across all commits containing each pattern and whether they
remain stable across projects. Including commits classified as unchanged, all four
patterns have negative median relative energy changes, ranging from $-2.72\%$ to
$-10.71\%$. However, the project-bootstrap 95\% confidence intervals for the
median remain entirely below zero only for \{Change Parameter Type, Inline
Variable\} and the complete triple; the intervals for the other two pairs include
zero. Association with a significant decrease therefore does not imply consistent
energy savings in every commit containing a pattern.

All four associations retain BH-corrected significance in 94.5\%--98.3\% of the
238 leave-one-project-out analyses. The triple also remains significant when its
comparison group is restricted to the 193 multi-type commits containing at least
three distinct refactoring types. Sensitivity to the minimum-support threshold is
less uniform: corrected decrease associations remain at thresholds from 3 to 11
commits, but the surviving patterns change, and none remains at thresholds from
12 to 15. Overall, the robustness analyses show that the decrease-associated
signal is not driven by any single project and remains detectable across multiple
analysis settings.

\paragraph{Overall refactoring patterns in practical commits.}
Taken together, the practical results show that neither individual refactoring
types nor the number of types in a commit alone adequately characterizes its
energy behavior. Although 256 of the 345 analyzed commits (74.2\%) contain
multiple refactoring types, multi-type commits are not significantly more likely
than single-type commits to exhibit an energy difference, and no individual type
shows a repeatable association after BH correction. By contrast, the
co-occurrence analysis identifies recurring pairs and triples associated with
significant energy decreases, and these associations remain detectable across
multiple robustness settings. This indicates that refactoring composition
provides useful information beyond refactoring multiplicity. These findings
complement the controlled results from \microbenchmark{}, which isolate an
intended refactoring type and reveal its possible energy behavior under specific
workloads. Practical commits combine multiple types and operations across files
and execute them under project-specific workloads; consequently, isolated
type-level effects cannot simply be transferred or added together to predict the
energy impact of a complete commit. Together, the two benchmarks show that the
energy impact of refactoring should be interpreted through both refactoring
composition and execution context. RQ3 therefore examines the source-level and
runtime factors that may explain why particular refactoring changes exhibit
energy differences.

\paragraph{Consistency of refactoring-type energy effects across benchmarks.}
To determine whether the workload-consistent effects observed in
\microbenchmark{} also appear in real-world commits, we compare the 18
\microbenchmark{} types that exhibit statistically significant differences in
the same direction across all six workload groups with the refactoring types
in the 89 single-type commits. Focusing on single-type commits
reduces the ambiguity caused by co-occurring refactoring types and provides the
clearest available practical evidence for comparison. Of these 18
\microbenchmark{} types, only Extract Class and Replace Loop with Pipeline also
occur among the 26 types in the single-type commits.

\begin{table}[t]
\centering
\caption{Cross-benchmark comparison of the consistently directional
\microbenchmark{} types that also occur among the single-type commits.
Practical outcomes are followed by their mean energy changes in parentheses.}
\label{tab:rq2_cross_benchmark_types}
\begin{adjustbox}{max width=0.8\columnwidth}
\begin{tabular}{lcccc}
\toprule
Refactoring type &
\microbenchmark{} direction &
Median $\Delta E$ &
Single-type commits &
Practical outcome \\
\midrule
Extract Class
& Increase (6/6) & +8.92\% & 1 & Unchanged (+5.64\%) \\
Replace Loop with Pipeline
& Increase (6/6) & +13.38\% & 1 & Unchanged (+15.17\%) \\
\bottomrule
\end{tabular}
\end{adjustbox}
\end{table}

As Table~\ref{tab:rq2_cross_benchmark_types} shows, both types significantly
increase energy consumption across all six workload groups in
\microbenchmark{}. Their single-type practical commits also have positive mean
energy changes, but neither difference is statistically significant. Moreover,
each type is represented by only one single-type commit, and none of the
consistently decreasing \microbenchmark{} types occurs among the single-type
commits. The practical evidence is therefore insufficient to confirm that
either type consistently increases energy across the two benchmarks. Overall,
our results do not identify any refactoring type that can currently be
classified as universally energy-increasing or energy-decreasing; establishing
such a classification would require repeated single-type observations across
diverse real-world projects.

\subsection{RQ3: What factors explain refactoring-induced energy differences?}
\label{subsec:result_rq3}

RQ1 established the frequency, direction, and magnitude of energy differences after refactoring, while RQ2 examined how these differences vary across refactoring types and combinations. RQ3 investigates the factors that help explain these differences. We first use \microbenchmark{} to relate refactoring-induced energy differences to changes in runtime behavior and examine their sensitivity to workload and JVM optimization. We then use \practicalbenchmark{} to examine which measured characteristics of real-world refactoring commits are associated with energy differences.

\subsubsection{Runtime Mechanisms in \microbenchmark{}}
\label{subsubsec:result_rq3_microbenchmark}

\paragraph{Analysis population and runtime measures.}
We analyze all \microbenchmark{} items for which both the energy measurements
and the corresponding runtime profiles are complete. Of the 71 candidate
items, 64 have valid energy measurements for Aggregate and all five input-size
workloads, as reported in RQ1. Among these items, one does not have a complete
JMH/JFR profile, leaving 63 items for the runtime analysis. This population
contains $63 \times 6 = 378$ item--workload observations and 11,340 paired
energy measurements. For every observation, we relate the mean relative energy
change to changes in execution time, normalized allocation per operation,
allocation samples, young-generation garbage-collection events, total
garbage-collection pause time, and CPU-load records. The runtime profiles are
collected separately from the energy measurements using the same before/after
versions and workload settings, thereby avoiding profiler overhead in the
energy measurements. Each runtime profile contains one JMH/JFR run per version;
we therefore interpret the runtime results as explanatory evidence rather than
as repeated performance estimates.

\begin{figure*}[t]
  \centering
  \includegraphics[width=0.7\textwidth]
  {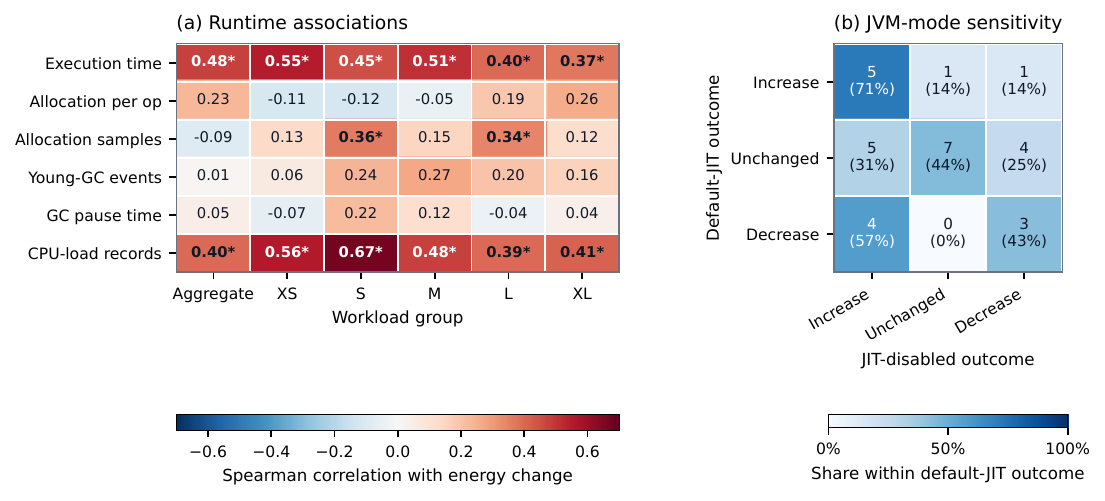}
  \caption{Runtime relationships and JVM-optimization sensitivity in
  \microbenchmark{}. (a) Spearman correlations between relative energy change
  and runtime-metric change for the 63 items with complete energy and JMH/JFR
  results. Asterisks denote associations that remain significant after
  Benjamini--Hochberg correction ($q<0.05$). Event-count and pause-time changes
  use signed log-transformed differences. (b) Transitions among significant
  increase, unchanged, and significant decrease outcomes from default-JIT
  execution (rows) to JIT-disabled execution using \texttt{-Xint} (columns) for
  the 30 items with complete Aggregate results under both modes. Cell labels
  report counts and row percentages.}
  \Description{Two heatmaps. The runtime-association heatmap shows consistent
  positive relationships for execution time and CPU-load records across all
  workload groups, while allocation and garbage-collection relationships are
  more selective. The JVM-mode heatmap shows that only half of the matched
  items retain the same energy-outcome category after JIT compilation is
  disabled.}
  \label{fig:rq3_micro_runtime_heatmaps}
\end{figure*}

Execution time and normalized allocation are expressed as relative changes.
For event and pause metrics that can equal zero, we use the signed difference
between their log-transformed values, $\log(1+X_{\mathrm{after}})-
\log(1+X_{\mathrm{before}})$, so that zero-to-nonzero transitions remain
defined. Because the change distributions are strongly skewed and include a
verified extreme item, Figure~\ref{fig:rq3_micro_runtime_heatmaps}(a) reports
Spearman correlations rather than correlations that can be dominated by a
single extreme value. We control the false-discovery rate across the 36
metric--workload associations using the Benjamini--Hochberg procedure.

\paragraph{Integrated relationships between runtime behavior and energy.}
Figure~\ref{fig:rq3_micro_runtime_heatmaps}(a) shows that execution-time change
has a positive association with energy change in every workload group. The
correlations range from $\rho=0.368$ for XL to $\rho=0.551$ for XS, and all six
remain significant after correction ($q\leq0.009$). CPU-load records show a
similar workload-wide pattern, with correlations from $\rho=0.394$ to
$\rho=0.673$ ($q\leq0.005$). Since these values count recorded CPU-load
events rather than directly measuring processor utilization, we treat them as
supporting rather than independent evidence. Among the 194 item--workload
observations with a statistically significant energy difference in this
63-item population,
execution time changes in the same direction as energy in 146 cases (75.3\%);
the workload-specific agreement ranges from 66.7\% for XL to 87.5\% for M. As
a consistency check using the energy-run records, mean batch duration agrees
in direction with energy in 182 of the 194 cases (93.8\%), compared with 114
cases (58.8\%) for mean power. Together, these results identify execution time as the most consistent measured pathway through which the controlled
transformations are associated with energy differences.

Allocation samples count sampled heap-object allocation events during execution and therefore reflect objects created by the program. Young-generation garbage-collection events count collections that mainly reclaim short-lived heap objects, while garbage-collection pause time measures the total time spent in garbage-collection pauses. Allocation per operation reports the average heap memory allocated for the workload. CPU-load records count the JFR CPU-load observations captured during execution. As shown in Figure~\ref{fig:rq3_micro_runtime_heatmaps}(a), changes in allocation
samples are positively associated with energy change under S
($\rho=0.361$, $q=0.010$) and L ($\rho=0.341$, $q=0.016$), but not under the
other four workloads. Normalized allocation per operation, young-generation
garbage-collection events, and garbage-collection pause time show no association
that remains significant after correction under any workload.

The absence of a benchmark-wide monotonic relationship does not imply that
memory allocation and garbage collection are irrelevant to individual
transformations. For example, as shown in
Table~\ref{tab:example_runtime_profile}, \textit{Replace Loop with Pipeline}
increases energy under all six workload groups, while the accompanying changes
in execution time, allocation samples, young-generation garbage-collection
events, and garbage-collection pause time vary substantially across workloads.
Thus, memory-allocation and garbage-collection activity can provide supporting
explanations for particular transformations and workloads, whereas execution
time provides the most consistent explanation across items. The remaining cases
in which execution time and energy change in different directions further show
that execution time alone cannot account for every observed energy difference.

\paragraph{Workload and JVM-optimization sensitivity.}
The runtime associations are workload-dependent. Execution-time change remains
associated with energy change across Aggregate and XS--XL, although its strength
varies, while allocation-sample change is associated only under S and L. Item-level
profiles further show that the magnitudes of execution-time, allocation, and
garbage-collection changes vary with input size. Thus, workload affects both whether an energy difference is detected and which runtime changes accompany it.

We assess JVM-optimization sensitivity by comparing Aggregate relative before/after
effects under default JIT and JIT-disabled execution with \texttt{-Xint}. Of the 64
RQ1-eligible items, 30 (46.9\%) yield 30 complete before/after energy pairs under
both modes. The remaining 34 do not yield complete valid \texttt{-Xint} results
within the measurement setup, which uses a 7,200-second timeout, and are excluded. Because the stored states do not distinguish timeouts from other
failures and the complete-case subset may favor faster items, these results apply
only to the 30 matched items. We compare within-item relative effects because batch
repetition counts can differ between JVM modes.

Figure~\ref{fig:rq3_micro_runtime_heatmaps}(b) shows substantial item-level
sensitivity. The default-JIT outcomes comprise seven increases, 16 unchanged cases,
and seven decreases, compared with 14 increases, eight unchanged cases, and eight
decreases under \texttt{-Xint}. Only 15 items (50.0\%) retain the same outcome
category ($\kappa=0.272$), and the signs of the raw change ratios agree for only 16
(53.3\%). Among the 14 items significant under default JIT, eight remain significant
in the same direction, five reverse direction, and one becomes unchanged; nine of
the 16 previously unchanged items become significant under \texttt{-Xint}.

The relative effects show no systematic shift between modes (Wilcoxon signed-rank
test, $p=0.221$) and no monotonic association in item rankings
($\rho=-0.001$, $p=0.997$). The latter result remains nonsignificant after excluding
\textit{Replace Error Code with Exception} ($\rho=-0.108$, $p=0.578$).
Disabling JIT therefore does not shift all refactorings in a common direction;
instead, it changes the direction, significance, or magnitude of many individual
effects.

\begin{findingbox}
Execution time and CPU-load records changes provide the most consistent
runtime-level explanations for refactoring-induced energy differences in
\microbenchmark{}: both are positively associated with energy change under every
workload. Execution time change also agrees in direction with 75.3\% of
statistically significant item--workload results. Allocation and
garbage-collection relationships are more workload-specific. JVM optimization
further changes the item-level outcome: only half of the 30 matched items retain
the same increase, unchanged, or decrease classification when JIT compilation
is disabled.
\end{findingbox}

\subsubsection{Explanatory Factors in \practicalbenchmark{}}
\label{subsubsec:result_rq3_practicalbenchmark}

\paragraph{Analysis population and factor families.}
We analyze the same 481 \practicalbenchmark{} commits from 430 projects used
in RQ1. For each commit, $\Delta E$ and $\Delta T$ denote the before/after log
ratios of energy and execution time. We define
$\Delta R=\Delta E-\Delta T$, where $R=E/T$ is the derived energy consumed per
unit of execution time rather than directly measured power. All commits have valid energy summaries and 30 paired before/after
execution-time observations. Static-code metrics (A) are available for 461
commits, code-change metrics (B) for 481, coverage metrics (C) for 418,
RefactoringMiner metrics (D) for 479, semantic replacement distance (E) for
454, and covered-diff alignment for 255. Each association uses the largest
valid population; factor-family comparisons use the 384 commits from 350
projects with complete execution-time and A--E values.

\begin{figure*}[t]
  \centering
  \includegraphics[width=0.8\textwidth]
  {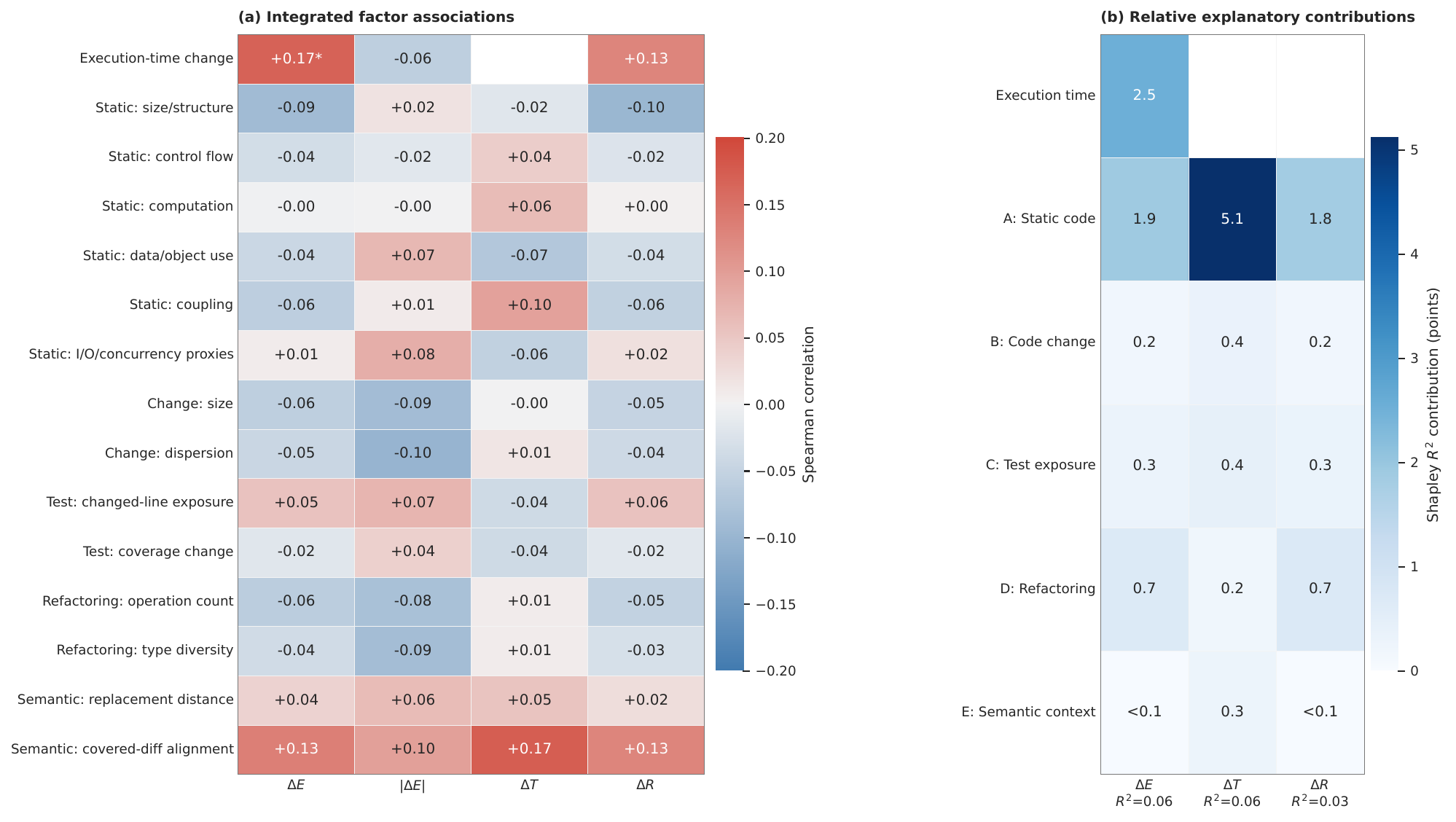}
  \caption{Explanatory-factor analysis for \practicalbenchmark{}.
  (a) Spearman correlations between execution and A--E factors and energy,
  energy-change magnitude, execution time, and derived energy-rate change.
  Each factor uses its largest valid population ($N=255$--$481$); the
  execution-time self-correlation is omitted. The asterisk marks the only
  association significant after BH correction across the 59 tests
  ($q<0.05$). (b) Shapley contributions to rank-based ordinary-least-squares
  $R^2$ for the common population of 384 commits from 350 projects. Cells
  report percentage-point contributions, and column labels report total
  $R^2$. The $\Delta E$ model includes execution time and A--E; the
  $\Delta T$ and $\Delta R$ models include A--E.}
  \Description{Two heatmaps showing weak associations and small explanatory
  contributions. Only execution-time change remains significantly associated
  with energy change after correction.}
  \label{fig:rq3_practical_factor_heatmaps}
\end{figure*}

We consolidate the 59 static metrics into six standardized, signed
log-transformed groups: size and structure, control flow, computation, data and
object use, coupling, and I/O or concurrency proxies. The remaining factors
capture change size and dispersion, changed-line coverage and its change,
refactoring count and diversity, and semantic replacement distance and
covered-diff alignment. Undefined static aggregates caused by absent constructs
are set to zero; extraction failures remain missing. We calculate Spearman
correlations with $\Delta E$, $|\Delta E|$, $\Delta T$, and $\Delta R$, apply
BH correction across the 59 associations in
Figure~\ref{fig:rq3_practical_factor_heatmaps}(a), and obtain confidence
intervals by resampling projects.

\paragraph{Integrated associations across execution and A--E factors.}
Figure~\ref{fig:rq3_practical_factor_heatmaps}(a) shows a weak positive
association between execution-time and energy changes
($\rho=0.167$, project-bootstrap 95\% CI $[0.084,0.247]$, $q=0.013$), the only
displayed association that remains significant after BH correction. Among the
32 commits with significant energy differences and with complete execution-time and A–E values, $\Delta E$ and $\Delta T$ have the same sign in 25 cases (78.1\%). However, execution time changes
significantly in the same direction in only six cases; it is unchanged in 25
and changes significantly in the opposite direction in one. Sign agreement
therefore overstates the explanatory role of execution time.

Execution-time changes are also small: the median
$|\Delta T|/|\Delta E|$ is 0.008, and $|\Delta T|$ reaches half of
$|\Delta E|$ in only two cases. In contrast, $\Delta R$ has the same direction
as $\Delta E$ and reaches at least half its magnitude in 31 cases, with a
median $|\Delta R|/|\Delta E|$ of 0.996. Thus, most significant energy
differences are reflected in the derived energy-rate change rather than
execution-time change. No individual A--E association remains significant after correction, and all coefficients satisfy $|\rho|\leq0.172$. Hence, none of the measured source, change, coverage, refactoring, or semantic factors provides a consistent monotonic explanation across commits, although commit-specific effects may
still exist.

\paragraph{Relative explanatory contributions.}
Figure~\ref{fig:rq3_practical_factor_heatmaps}(b) shows that execution time
contributes 2.5 $R^2$ percentage points to $\Delta E$, or 44.6\% of the
complete model's explained variation. Static-code and refactoring factors
contribute 1.9 and 0.7 points, respectively; each remaining family contributes
at most 0.3 points. Nevertheless, the complete model explains only 5.7\% of
the rank variation in $\Delta E$. The A--E factors explain 6.4\% of the
variation in $\Delta T$ and 3.1\% in $\Delta R$, with static-code factors
providing most of these limited contributions. Most between-commit variation
therefore remains unexplained.

\begin{findingbox}
In \practicalbenchmark{}, execution-time change is the only measured factor
that remains associated with energy change after correction, but the
relationship is weak, and a significant timing change agrees
with the energy outcome in only 6 of the 32 commits with a significant energy
difference. No individual static-code, code-change, test-exposure,
refactoring, or semantic factor remains significant, and all measured factors
together explain only 5.7\% of the rank variation in energy change. Thus,
execution time and the available commit-level characteristics do not
provide a sufficient general explanation for refactoring-induced energy
differences in real-world commits.
\end{findingbox}

\paragraph{Overall explanation across the two benchmarks.}
Across both benchmarks, behavior preservation does not guarantee identical
runtime behavior or energy consumption. In \microbenchmark{}, execution time
provides the most consistent explanation, while allocation and
garbage-collection relationships depend on workload and many outcomes change
when JIT compilation is disabled. In \practicalbenchmark{}, execution time
remains the strongest measured factor, but its association is weak and the
available commit-level factors explain little variation. This difference reflects the heterogeneous transformations and execution
contexts of practical commits, whose static characteristics only indirectly
represent runtime behavior. Overall, refactoring-induced energy differences
arise from interactions among the transformation, workload, and JVM behavior,
rather than from fixed refactoring-type effects or coarse code characteristics.
The limited explanatory power of these factors motivates RQ4, which examines
whether they can nevertheless identify energy regressions.

\subsection{RQ4: How effective are existing techniques in identifying refactoring-induced energy regressions?}
\label{subsec:result_rq4}

To answer \rqfour{}, we evaluate two approaches on
\practicalbenchmark{}: metric-based classifiers adapted from prior
performance-bug prediction work~\cite{10.1145/3643991.3644920} and a
prompt-based LLM classifier. Following the ground-truth definition in RQ1, a
commit is labeled \textsc{Reg} when the after-refactoring version exhibits a
statistically significant energy increase under the paired Wilcoxon
signed-rank test ($p<0.05$) and has a positive mean paired energy difference.
All other commits, including those with no significant difference and those
with a significant decrease, are labeled \textsc{Not Reg}. We use precision,
recall, and F1 for \textsc{Reg} as the primary measures of identification
effectiveness.

\paragraph{Origin and adaptation of the existing techniques.}
Our metric-based baselines are adapted from the performance-bug prediction
approach of~\citet{10.1145/3643991.3644920}. Their study predicts
performance regression using code metrics and
evaluates seven machine-learning algorithms. We select the three
best-performing algorithms reported in their study: Logistic Regression,
Random Forest, and XGBoost. For our task, each metric-based classifier uses
all five information families: static-code metrics (A), code-change metrics
(B), test/coverage metrics (C), refactoring metrics (D), and code-diff
information (E). We additionally evaluate \texttt{gpt-5-mini} using the same five information families (A--E).

\paragraph{Metric-based identification.}
Table~\ref{tab:rq4_approach_comparison} marks the three classifiers adapted
from Zhao et al.~\cite{10.1145/3643991.3644920} and reports their performance
using A--E. Logistic Regression is the only metric-based classifier that
identifies any regression cases, but its precision of 0.031 and recall of
0.200 yield an F1 of only 0.053. Random Forest and XGBoost obtain zero
precision, recall, and F1 for \textsc{Reg}, meaning that neither identifies a
true regression under the evaluated setting.

\begin{table}[t]
  \centering
  \caption{Performance on the \textsc{Reg} class across identification
  techniques.}
  \label{tab:rq4_approach_comparison}
  \begin{adjustbox}{width=0.4\linewidth}
    \begin{tabular}{lccc}
      \toprule
      \textbf{Tool} & \textbf{Precision} & \textbf{Recall} & \textbf{F1} \\
      \midrule
      \texttt{gpt-5-mini}
        & \textbf{0.056} & \textbf{0.357} & \textbf{0.096} \\
      Logistic Regression\textsuperscript{*}
        & 0.031 & 0.200 & 0.053 \\
      Random Forest\textsuperscript{*}
        & 0.000 & 0.000 & 0.000 \\
      XGBoost\textsuperscript{*}
        & 0.000 & 0.000 & 0.000 \\
      \bottomrule
    \end{tabular}
  \end{adjustbox}

  \vspace{2pt}
  \begin{minipage}{0.98\linewidth}
    \footnotesize
    \textsuperscript{*}Existing metric-based technique adapted from Zhao et
    al.~\cite{10.1145/3643991.3644920}. These are the three best-performing
    algorithms in their study.
  \end{minipage}
\end{table}

\paragraph{LLM-based identification.}
Table~\ref{tab:rq4_llm_ablation} reports the performance of
\texttt{gpt-5-mini} when its structured prompt contains all five information
families and when one family is omitted at a time. With all families included, the model achieves a \textsc{Reg} precision of 0.056, recall of 0.357, and F1 of
0.096. Thus, it identifies 35.7\% of the actual regressions but misses the
remaining 64.3\%. Moreover, only 5.6\% of its regression predictions are
correct, equivalent to approximately one true regression for every 18
commits flagged for validation.

\begin{table*}[t]
  \centering
  \caption{Performance of the LLM-based technique and its
  leave-one-information-family-out ablations.}
  \label{tab:rq4_llm_ablation}
  \begin{adjustbox}{width=0.5\textwidth}
    \begin{tabular}{llccc}
      \toprule
      \multirow{2}{*}{\textbf{Tool}} &
      \multirow{2}{*}{\textbf{Omitted information}} &
      \multicolumn{3}{c}{\textbf{\textsc{Reg}}} \\
      \cmidrule(lr){3-5}
      & & \textbf{Precision} & \textbf{Recall} & \textbf{F1} \\
      \midrule
      \multirow{6}{*}{\texttt{gpt-5-mini}} & None
        & \textbf{0.056} & \textbf{0.357} & \textbf{0.096} \\
      & Static-code metrics (A)$^\dagger$
        & 0.040 & 0.286 & 0.070 \\
      & Code-change metrics (B)$^\dagger$
        & 0.031 & 0.214 & 0.055 \\
      & Test/coverage metrics (C)
        & 0.011 & 0.071 & 0.019 \\
      & Refactoring metrics (D)
        & 0.040 & 0.286 & 0.070 \\
      & Code diff (E)
        & 0.024 & 0.143 & 0.041 \\
      \bottomrule
    \end{tabular}
  \end{adjustbox}

  \vspace{2pt}
  \begin{minipage}{0.98\textwidth}
    \footnotesize
    Each row after the first omits the indicated information family from the
    complete prompt. $^\dagger$Information family adapted from the existing
    metric-based technique of Zhao et
    al.~\cite{10.1145/3643991.3644920}; C--E are additional information used
    in this study. For E, the LLM receives the textual code diff. Bold
    values are the highest in each reported column.
  \end{minipage}
\end{table*}

\paragraph{Contribution of the information families.}
The complete A--E prompt obtains the highest \textsc{Reg} precision, recall,
and F1, indicating that no individual family can be removed without reducing
target-class performance. Test/coverage metrics (C) have the largest observed
ablation effect: omitting them reduces \textsc{Reg} recall from 0.357 to
0.071 and \textsc{Reg} F1 from 0.096 to 0.019. Code-diff information (E)
provides the next clearest minority-class signal; its removal reduces
\textsc{Reg} recall to 0.143 and F1 to 0.041. Removing code-change metrics (B)
also lowers recall to 0.214, whereas omitting static-code (A) or refactoring
(D) metrics lowers it to 0.286. These leave-one-family-out results suggest
that test exposure and the textual content of the code change provide
complementary evidence to the LLM.

\paragraph{Cross-approach comparison.}
Table~\ref{tab:rq4_approach_comparison} compares the complete LLM configuration with the three metric-based classifiers on the \textsc{Reg} measures. The LLM has the highest reported values on all three measures: compared with Logistic Regression, its recall is 0.357 rather than 0.200 and its F1 is 0.096 rather than 0.053. More importantly, the LLM's numerical advantage does not make its absolute performance adequate. It still misses almost two thirds of the regressions and produces predominantly false regression alerts. The uniformly low absolute results show that neither the adapted existing techniques nor the LLM-based technique provides a dependable decision boundary for this rare outcome.

\begin{findingbox}
Metric-based techniques adapted from prior performance-bug prediction transfer
poorly to refactoring-induced energy-regression identification: Logistic
Regression achieves only 0.053 F1, while Random Forest and XGBoost identify no
true regressions. Using all five information families, \texttt{gpt-5-mini}
performs best among the evaluated configurations, but its 0.357 recall and
0.056 precision mean that it misses 64.3\% of regressions and yields only about
one true regression per 18 alerts. Consequently, none of the evaluated
techniques is reliable as a standalone replacement for direct energy
validation.
\end{findingbox}
\section{Implications}
\label{sec:implication}

Refactoring preserves program behaviour but does not guarantee preservation of energy consumption. Across both controlled and real-world benchmarks, refactoring can increase, decrease, or leave energy consumption unchanged, depending on the transformation, workload, and execution context. These findings show that the energy impact of refactoring is context-dependent rather than deterministic, leading to several implications.

\paragraph{Developers should validate energy-sensitive refactorings under
representative workloads.}
Knuth argues that programmers should avoid optimizing noncritical parts of a
program~\cite{10.1145/356635.356640}. Our results lead to the same practical
lesson for energy: developers should first identify the code and workloads that
materially contribute to energy consumption, and then measure the refactorings
that affect them. They should compare the original and refactored versions
through repeated paired measurements under representative workloads. Workload
selection is crucial: in \microbenchmark{}, 45.3\% of items change
classification across input sizes, and 15.6\% switch between increases and
decreases.

\paragraph{Energy-aware refactoring tools should verify predictions with
targeted measurement.}
Execution time change is the clearest measured mechanism in
\microbenchmark{}, but it is only weakly associated with energy change in
\practicalbenchmark{}. The best evaluated LLM also misses 64.3\% of regressions,
and only about one in 18 of its regression alerts is correct. No evaluated
proxy or predictor is therefore reliable enough to replace energy measurement.
Moreover, tools should not directly extrapolate isolated type-level effects to
real-world commits, particularly when multiple refactorings may interact or
offset one another. Tools can instead use refactoring, code change, coverage,
and runtime evidence to select candidates and representative workloads, then
confirm consequential cases through paired measurements. This keeps
measurement in the decision loop while using prediction to reduce its cost.

\paragraph{Researchers should study refactoring as an execution-conditioned
change.}
Our results suggest that a refactoring type is too coarse to serve as a
universal energy label. Future benchmarks should treat the combination of
transformation pattern, workload, and runtime environment as the unit of
analysis, with repeated measurements and diverse workloads.  They should also
analyze single-type and mixed-type commits separately. This would reveal when a
strong isolated effect matters in practice and when it becomes small or is
offset by other refactorings.

\section{Threats to Validity}
\label{sec:threats_to_validity}

\noindent\textbf{Internal.}
Energy measurements can be affected by JVM warm-up, background
activity, and execution workloads. We mitigate this threat by conducting all
experiments on a dedicated machine without concurrent workloads, warming up the
JVM before measurement, inserting cooling intervals between runs, and
randomizing the execution order of the before- and after-refactoring versions.
Within each pair, we keep the workload and execution configuration fixed. In
\microbenchmark{}, we additionally use a common repetition count and collect 30
paired measurements. We use the paired Wilcoxon signed-rank test and report
Cliff's delta to reduce sensitivity to measurement noise and distributional
assumptions. In \practicalbenchmark{}, non-refactoring changes within a commit could confound
the observed energy difference. We reduce this threat by requiring both versions
to compile and pass the same unchanged tests, excluding commits that modify test
files, requiring positive modified-line coverage measured by JaCoCo, and
manually inspecting the retained commits for non-refactoring changes. These
controls reduce the influence of broken builds, changed test behavior, and
unexecuted or unrelated code on the measured differences.

\noindent\textbf{External.}
The measured energy impact of refactoring depends on the workload used to
exercise the program. We address this threat in \microbenchmark{} by evaluating
each eligible item under an aggregate workload and five input sizes. In
\practicalbenchmark{}, we repeatedly execute the test suite of each real
project. Requiring modified-line coverage further ensures that the changed code
is exercised during measurement. We also use two complementary benchmarks:
\microbenchmark{} isolates refactoring effects under controlled workloads,
whereas \practicalbenchmark{} captures refactoring in real repositories. This
combination provides evidence from both controlled and practical settings.

Our analysis focuses on Java programs. We chose Java because it is one of the
most widely used programming languages~\cite{10.1145/3459955.3460614}, has a
mature refactoring and execution ecosystem, and provides continuity with closely
related research. Fowler's original refactoring catalog uses Java
examples~\cite{fowler1999refactoring}; Sahin et al.\ study the energy impact of
refactorings on Java applications~\cite{10.1145/2652524.2652538}; and prior
performance-bug and performance-regression studies also examine Java
systems~\cite{10.1145/3643991.3644920,9197704,10.1145/3377811.3380351}.
This ecosystem enables \microbenchmark{} to cover 68 refactoring types and
allows real Maven projects in \practicalbenchmark{} to be compiled, tested, and
measured consistently.
\section{Related Work}
\label{sec:relatedwork}

\subsection{Energy Effects of Program Transformations}

Software energy consumption can change even when functionality is preserved.
\citet{6224257} show that the energy effects of design patterns vary across
individual patterns and cannot be inferred from broad pattern categories.
\citet{10.1145/2568225.2568297} present SEEDS, which searches among
behavior-preserving implementation alternatives and measures them under a
target workload to identify lower-energy choices. Other studies report that
code obfuscation can alter energy consumption~\cite{6976079} and propose
source-level transformations for removing energy-inefficient constructs in
embedded software~\cite{kim2018code}. Refactoring-specific studies provide the closest basis for our work.
\citet{10.1145/2652524.2652538} manually apply six refactoring types to nine
Java applications and measure energy through repeated JUnit-test executions.
They find both increases and decreases, and show that execution time and
dynamic execution counts do not fully predict the resulting energy effect.
\citet{park2014investigation} estimate the energy consumption of small examples
covering 63 Fowler refactoring types, whereas
\citet{csanlialp2022energy} examine triple combinations of five refactoring
types. Complementing these controlled energy studies,
\citet{10.1145/3485136} mine refactoring-focused commits and show that
refactoring can also produce execution-time improvements and regressions in
real Java systems. However, prior energy studies primarily examine isolated
examples under fixed or simple workloads. They therefore leave open how refactoring
energy effects vary across workloads, manifest in real-world commits, and relate to source-level and runtime factors.

\subsection{Performance-Regression Identification}

Prior work has used tests and software metrics to identify performance
regressions. \citet{10.1145/3377811.3380351} investigate whether tests already
available in release pipelines can serve as performance tests, and
\citet{9197704} propose test-level just-in-time prediction of
performance-regression-inducing commits. 
\citet{liao2020using,liao2021locating} leverage black-box machine learning models that learn performance from execution events to detect performance regressions.
More recently,
\citet{10.1145/3643991.3644920} predict performance regression using code metrics with multiple machine-learning algorithms. These techniques target execution-performance bugs or regressions,
not energy increases introduced by behavior-preserving refactoring. Our study connects these two research lines. We combine a controlled,
workload-aware \microbenchmark{} spanning 68 refactoring types with a
\practicalbenchmark{} of real-world GitHub refactoring commits, where multiple
operations may co-occur. Beyond measuring whether energy changes, we analyze
workload sensitivity, refactoring types and co-occurrence patterns, and
source-level and runtime factors associated with the changes. We further
evaluate whether metric-based techniques adapted from performance-bug
prediction and a prompt-based LLM can identify refactoring-induced energy
regressions. Thus, our study extends prior evidence from whether refactoring
can affect energy to when such effects arise, how they can be explained, and how effectively regressions can be identified.

\section{Conclusion}
\label{sec:conclusion}

This paper examined the energy impact of behavior-preserving refactoring using
a workload-aware \microbenchmark{} covering 68 refactoring types and a
\practicalbenchmark{} of 481 real-world refactoring commits. In \microbenchmark{}, 51.8\% of the 384 refactoring-instance--workload comparisons exhibit statistically
significant energy differences, and 45.3\% of the 64 eligible refactoring
instances yield different increase, decrease, or unchanged outcomes across input
sizes. In \practicalbenchmark{}, significant differences occur in only 7.5\% of
commits, although two thirds of these statistically significant changes have an
absolute magnitude of at least 10\%. Both benchmarks contain increases and
decreases. Energy effects cannot be assigned as fixed labels to refactoring types. In the controlled benchmark, the effects of individual refactoring types are often
workload-dependent; in \practicalbenchmark{}, no individual refactoring type has
a consistent association with energy change, although several recurring
combinations are associated with decreases. Across the \microbenchmark{}
workloads, energy change is associated with changes in execution time and
CPU-load-record count. In \practicalbenchmark{}, however, the execution-time relationship is weak, and all measured factors together explain only 5.7\% of the variation in energy change. Overall, behavior-preserving refactoring is not inherently
energy-neutral: its observed energy impact varies across refactoring types and
workloads and differs between isolated refactoring instances and real-world
refactoring commits. Neither the adapted metric-based classifiers nor the
evaluated LLM can reliably identify refactoring-induced energy regressions.

\section*{Acknowledgment}
We acknowledge the support of the Natural Sciences and Engineering Research Council of Canada (NSERC), [funding reference number ALLRP 597968 - 24].
Cette recherche a été financée par le Conseil de recherches en sciences naturelles et en génie du Canada (CRSNG), [numéro de référence ALLRP 597968 - 24].
Ce projet de recherche \# 361973 est rendu possible grâce au financement du Fonds de recherche du Québec. We acknowledge the support of the Government of Canada's New Frontiers in Research Fund (NFRF) under Grant No. NFRFE-2024-00612.
\bibliographystyle{ACM-Reference-Format}
\bibliography{ref}
\end{document}